%% file: gaiafpr.tex
\documentclass[apj,twocolumn,twocolappendix,floatfix]{openjournal}

\DeclareRobustCommand{\VAN}[3]{#2}
\let\VANthebibliography\thebibliography
\def\thebibliography{\DeclareRobustCommand{\VAN}[3]{##3}\VANthebibliography}

\usepackage{amsmath}	% Advanced maths commands
\usepackage{amssymb}	% Extra maths symbols
\usepackage{booktabs}
\usepackage[T1]{fontenc}
\usepackage{graphicx}	% Including figure files
\usepackage[breaklinks,colorlinks,citecolor=blue,urlcolor=blue,linkcolor=blue]{hyperref}
\usepackage{cleveref}
\usepackage{multirow}
\usepackage{newtxtext,newtxmath}
\usepackage{orcidlink}
\usepackage{pifont}% http://ctan.org/pkg/pifont
\usepackage[nobottomtitles]{titlesec}
\titleformat{\section}{\filcenter\MakeUppercase}{\thesection.}{0.5em}{}

\usepackage{savesym}
\savesymbol{tablenum}
\usepackage{siunitx}
\restoresymbol{SIX}{tablenum}

\newcommand{\PHOEBE}{\texttt{PHOEBE}}
\newcommand{\Gaia}{\textit{Gaia}}
\newcommand{\TESS}{\textit{TESS}}
\newcommand{\WISE}{\textit{WISE}}

\newcommand{\cmark}{\ding{51}}%
\newcommand{\xmark}{\ding{55}}%

\newcommand{\nFPRtotal}{9,614}
\newcommand{\nELLtotal}{1,109}
\newcommand{\nHighFM}{8}

\begin{document}
% Title of the paper, and the short title which is used in the headers.
% Keep the title short and informative.
\title[Gaia FPR ELLs]{High mass function ellipsoidal variables in the Gaia Focused Product Release: searching for black hole candidates in the binary zoo}

% The list of authors, and the short list which is used in the headers.
% If you need two or more lines of authors, add an extra line using \newauthor

\author{\vspace{-1.3cm}D. M. Rowan\,\orcidlink{0000-0003-2431-981X}$^{1,2}$}
\author{Todd A. Thompson\,\orcidlink{0000-0003-2377-9574}$^{1,2,3}$}
\author{T. Jayasinghe\,\orcidlink{0000-0002-6244-477X}$^{4}$}
\author{C. S. Kochanek\ $^{1,2}$}
\author{K. Z. Stanek\ $^{1,2}$}

\affiliation{$^{1}$Department of Astronomy, The Ohio State University, 140 West 18th Avenue, Columbus, OH, 43210, USA}
\affiliation{$^{2}$Center for Cosmology and Astroparticle Physics, The Ohio State University, 191 W. Woodruff Avenue, Columbus, OH, 43210, USA}
\affiliation{$^{3}$Department of Physics, The Ohio State University, Columbus, Ohio, 43210, USA}
\affiliation{$^{4}$Independent Researcher, San Jose, California, USA}

% Abstract of the paper
\begin{abstract}
The recent \Gaia{} Focused Product Release contains radial velocity time-series for more than 9,000 \Gaia{} long-period photometric variables. Here we search for binary systems with large radial velocity amplitudes to identify candidates with massive, unseen companions. Eight targets have binary mass function $f(M)>1\ M_\odot$, three of which are eclipsing binaries. The remaining five show evidence of ellipsoidal modulations. We fit spectroscopic orbit models to the \Gaia{} radial velocities, and fit the spectral energy distributions of three targets. For the two systems most likely to host dark companions, J0946 and J1640, we use \PHOEBE{} to fit the ASAS-SN light curves and \Gaia{} radial velocities. The derived companion masses are $>3\ M_\odot$, but the high Galactic dust extinctions towards these objects limit our ability to rule out main sequence companions or subgiants hotter than the photometric primaries. These systems are similar to other stellar-mass black hole impostors, notably the Unicorn (V723 Mon) and the Giraffe (2M04123153$+$6738486). While it is possible that J1640 and J0946 are similar examples of stripped giant star binaries, high-resolution spectra can be used to determine the nature of their companions.
\end{abstract}

\maketitle

\section{Introduction}

The mass distribution of stellar mass black holes (BHs) is directly related to the late-stage evolution of massive stars and core-collapse supernovae. By observing and characterizing the black hole population in the Milky Way, we stand to learn more about the initial-final mass relation (IFMR), which describes the connection between the pre-supernova mass of the star and the type of compact object produced. Various physical dependencies complicate the modeling of the IFMR, such as metallicity, mass-loss rate, and binary interactions \citep{Sukhbold16}. An accurate census of compact objects is also needed to understand the ``lower mass gap'' between the most massive neutron stars \citep[${\sim}2.1~M_\odot$,][]{Antoniadis16, Cromartie20} and the least massive BHs in X-ray binaries \citep[${\sim}5~M_\odot$,][]{Ozel10, Farr11, Kochanek15}.

The Milky Way is expected to contain ${\sim}10^8$ stellar-mass BHs and ${\sim}10^9$ neutron stars  \citep{Timmes96,Wiktorowicz19}. The compact objects detected so far are dominated by X-ray binaries \citep{Neumann23} and extra-Galactic gravitational wave mergers \citep[e.g.,][]{Abbott16}. However, only a small number of Galactic BHs are expected to be in the tight binary orbits necessary to produce X-ray emission \citep{Corral16}, and the fraction of systems that merge in gravitational wave events is also a strongly biased fraction of the overall population \citep[e.g.,][]{Kruckow18}. The majority of Galactic BHs are instead most likely in non-interacting binaries that are not \mbox{X-ray} bright or are isolated free-floating systems. Discovering and characterizing these systems is crucial to understanding the population and improving models for the end states of stellar evolution. While isolated BHs can only be detected via microlensing surveys \citep{Lam20, Lam22, Sahu22}, there are multiple methods that can be used to search for and characterize non-interacting BHs in binary systems. 

\Gaia{} astrometry has proven to be an effective tool to probe binary orbits. Population synthesis studies predict that \Gaia{} astrometry will detect tens to thousands of BHs \citep{Breivik17, Yamaguchi18}, though these predictions require assumptions about complicated physical processes like common envelope evolution and neutron star/BH natal kicks. So far, only two strong BH candidates have been identified using \Gaia{} astrometry (\Gaia{} BH-1: \citealt{ElBadry23_BH1, Chakrabarti23}; \Gaia{} BH-2: \citealt{Tanikawa23, ElBadry23_BH2}). The formation and evolutionary history of these systems are unclear, and there is some evidence that they formed from dynamical interactions in cluster environments \citep[e.g.,][]{Rastello23}.

Spectroscopic surveys can be used to search for single-lined binaries (SB1s) with large amplitude periodic radial velocity (RV) variations. Systems identified through RVs can only yield a lower limit to the companion mass because, unlike astrometric binaries, the orbital inclination is not directly constrained. For example, \citet{Thompson19} reported the detection of a possible stellar-mass black hole orbiting the spotted red giant 2MASS J05215658+4359220 using RVs from the Apache Point Observatory Galactic Evolution Experiment \citep[APOGEE,][]{Majewski17}. Some candidates have also been identified using spectroscopic data from the Large Sky Area Multi-object Fiber Spectroscopic Telescope \citep[LAMOST,][]{Cui12, Gu19}, but additional observations are needed to confirm their orbits and to rule out luminous companions. More recently, \Gaia{} Data Release 3 included spectroscopic orbits for $>180,000$~SB1s \citep{Gaia23_summary, Gaia23_hiddentreasure}, however no strong candidates have been identified \citep{ElBadry22, Jayasinghe23}. Aside from large spectroscopic surveys, dedicated observations of targets in globular clusters have identified several candidates \citep{Giesers18, Giesers19}, but dynamical effects in clusters may drive the formation and evolution of these systems, making comparison to the field population challenging \citep[e.g.,][]{Ryu23}. 

For some spectroscopic binaries, photometric ellipsoidal variability can be used to place additional constraints on the mass ratio and inclination of the binary. Ellipsoidal modulations are caused by the tidal distortion of a star by its binary companion. Some X-ray binary systems are ellipsoidal variables \citep[ELLs; e.g.,][]{Orosz01}, but the accretion and its variability can limit our ability to use ellipsoidal modulations to characterize these systems. There have been a number of searches to identify ELLs with high mass companions in photometric surveys \citep[e.g.,][]{Rowan21, Gomel21, Gomel23, Green23}, but spectroscopic follow-up is required for these systems and no strong candidates have been identified \citep[e.g.,][]{Nagarajan23}.

The presence of ellipsoidal modulations can also indicate a history of mass transfer that complicates the characterization of systems. For example, V723~Mon was initially reported as a red giant with a mass-gap black hole companion $M\sim 3\ M_\odot$ \citep{Jayasinghe21_unicorn}. The light curve shows clear ellipsoidal variability consistent with the RV orbital period and no strong evidence of a luminous companion in the spectra. Using spectral disentangling, \citet{ElBadry22_zoo} showed that the system is instead better described by a very low mass stripped red giant $M\approx 0.4\ M_\odot$ with a subgiant companion at a similar effective temperature. The subgiant is also rapidly rotating, blurring its spectral features. 2M04123153$+$6738486 was similarly identified as a BH false-positive better explained by a stripped star scenario \citep{Jayasinghe22}, demonstrating the complexities in ruling out luminous companions for ellipsoidal variables with large amplitude RV variations. 

As astrometric, spectroscopic, and photometric surveys expand, it is becoming increasingly important to be able to promptly and accurately vet targets and identify the strongest non-interacting BH candidates. \Gaia{} DR3 includes more than 800,000 non-single star solutions. The next \Gaia{} data release, not expected before 2025, will include many more binary star orbits as well as epoch RVs and astrometric measurements. The recent \Gaia{} Focused Product Release included time-series RV measurements for a sample of $>9,000$ stars identified as long-period variables (LPVs) in \Gaia{} photometry \citep{Gaia2023_FPR}. Here, we use this catalog to identify ellipsoidal variables with high radial velocity amplitudes that could host compact object companions.

In Section \S\ref{sec:target_selection}, we describe how we select targets with high binary mass function and reject obvious eclipsing binaries. We fit spectroscopic orbit models in Section \S\ref{sec:rvs} and spectral energy distributions (SEDs) in Section \S\ref{sec:seds}. Using the SED fit results, we model the light curve and radial velocity curve using PHysics Of Eclipsing BinariEs \citep[\PHOEBE{},][]{Prsa05, Conroy20} in Section \S\ref{sec:phoebe}. Finally, we present possible interpretations of the systems and describe the additional observations that are necessary to better characterize these binaries in Section \S\ref{sec:discussion}.

\section{Target Selection}\label{sec:target_selection}

\begin{table*}
    \centering
    \caption{High mass function binary candidates in the \Gaia{} FPR sample. The period, $P$, and radial velocity semiamplitude, $K$, are from \citet{Gaia2023_FPR}. The distance column gives the median photogeometric distance from \citet{BailerJones21}.}
    \sisetup{table-auto-round,
     group-digits=false}
    \setlength{\tabcolsep}{10pt}
    \begin{center}
        \input{ANC/target_selection_table}
    \end{center}
    \label{tab:target_selection}
\end{table*}

\begin{figure}
    \centering
    \includegraphics[width=\linewidth]{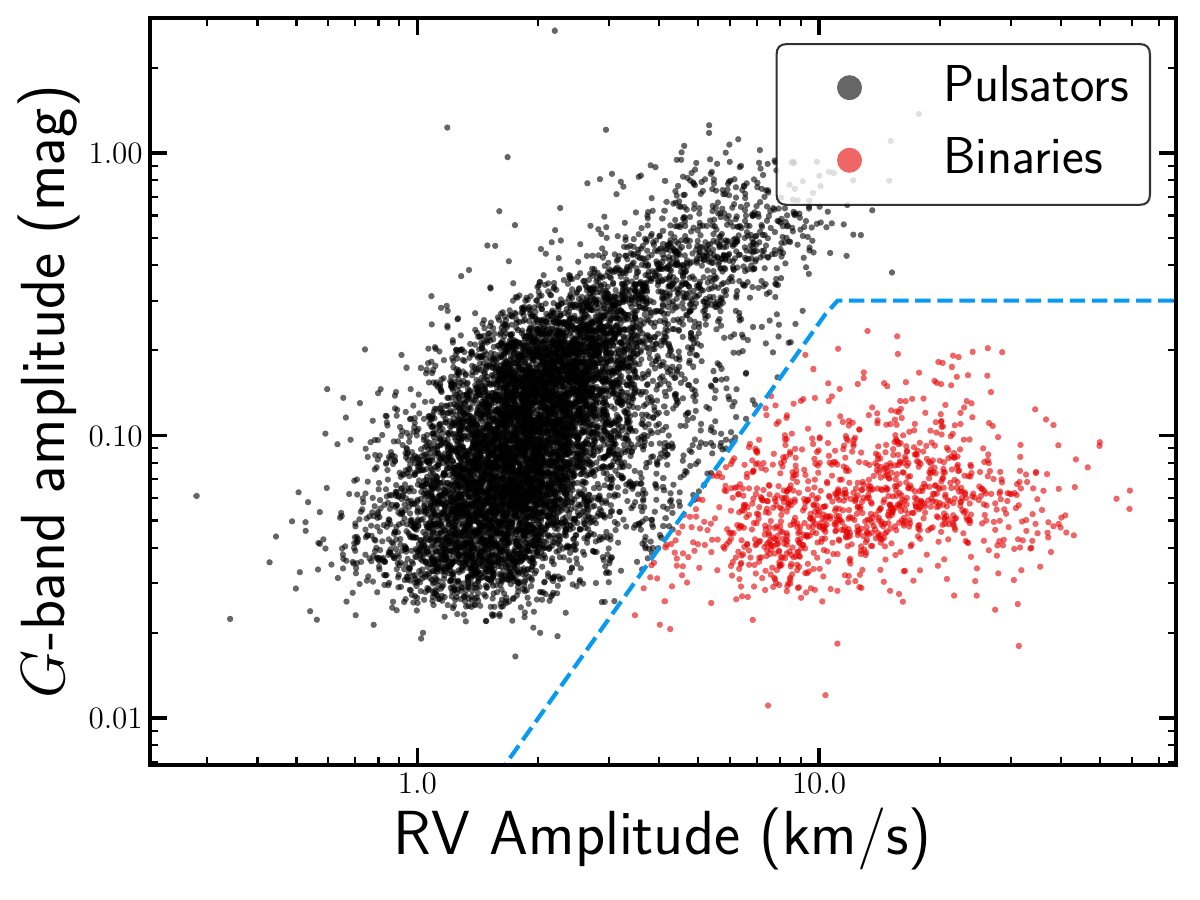}
    \caption{The amplitude of photometric and radial velocity variability for the full \Gaia{} FPR LPV sample. The blue line shows the cut described by \citet{Gaia2023_FPR} that separates binary stars from pulsating variables.}
    \label{fig:target_selection}
\end{figure}

\begin{figure}
    \centering
    \includegraphics[width=\linewidth]{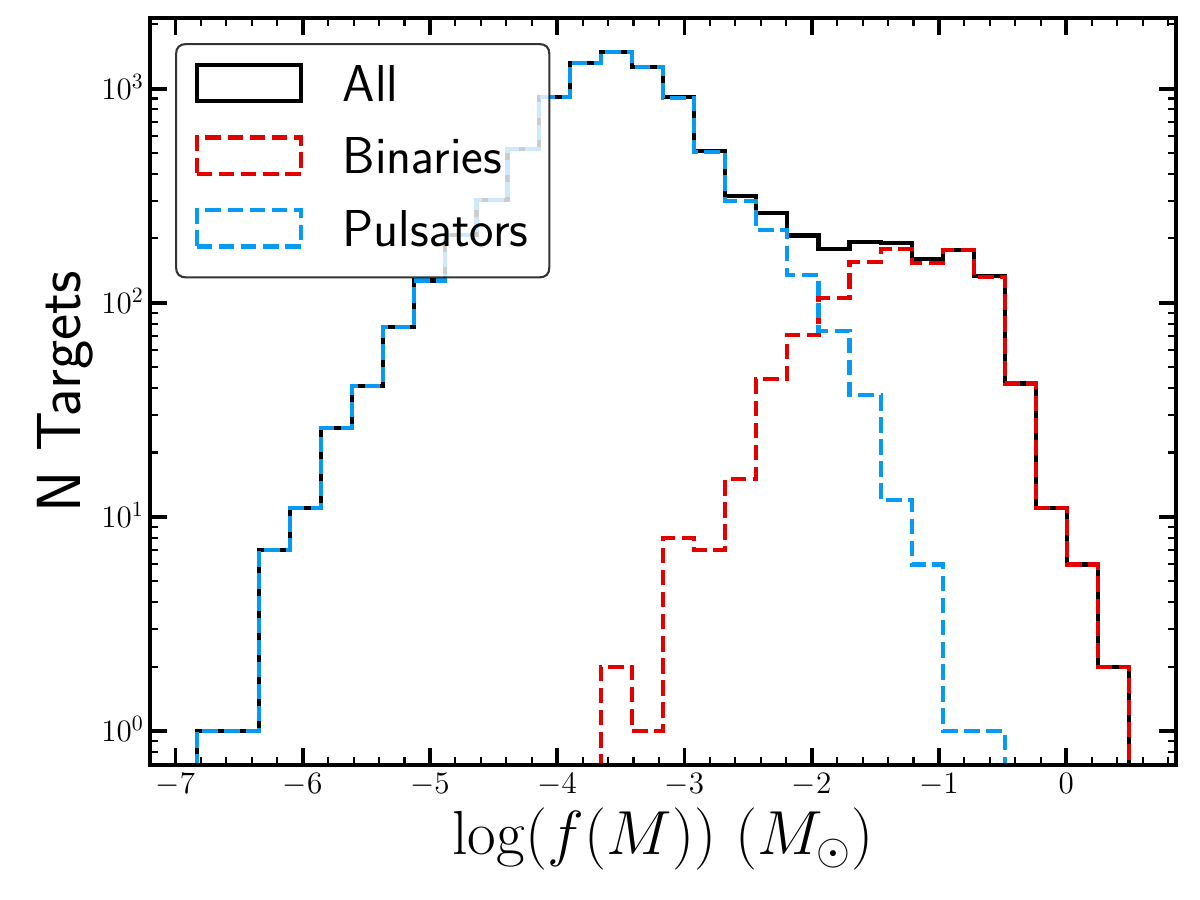}
    \caption{Distribution of the binary mass function for the full sample of targets in the \Gaia{} FPR catalog. There are \nHighFM{} targets with $f(M) > 1.0\ M_\odot$, and all are binaries based on the cut from Figure \ref{fig:target_selection}.}
    \label{fig:fm_dist}
\end{figure}

The details of the \Gaia{} FPR catalog construction are described in \citet{Gaia2023_FPR}. In brief, the process starts from the full sample of more than 1.7 million \Gaia{} long-period variables \citep{Lebzelter23}. Only $\sim500,000$ of these have \Gaia{} RV measurements. After making cuts on the magnitude ($G_{\rm{RVS}}<12$~mag), the number of visibility periods ($>12$), and on the ratio $\epsilon_{\rm{RV}}/A_{\rm{RV}} < 0.175$ between the RV amplitude, $A_{\rm{RV}}$, and the RV uncertainty $\epsilon_{\rm{RV}}$, only $\sim111,000$ targets remain. The photometric and RV time-series are then searched for periodic signals. Systems with long-term linear RV trends trends are also removed leaving only $\sim 44,000$ systems. Finally, targets are selected to have photometric periods equal to or twice the best-fit period of the RV time series. The final sample contains \nFPRtotal{} long-period variables with epoch RVs. This is the second sample of epoch RVs from \Gaia{}, following the samples of $\sim2,000$ Cepheids \citep{Ripepi23} and RR Lyrae variables \citep{Clementini23} included in \Gaia{} DR3.

\Gaia{} DR3 also includes a large number of binary star orbits as part of the ``non-single-stars'' (NSS) tables. We cross-match the \Gaia{} FPR catalogs with the Gaia NSS tables and find a total of 862 matches, 854 of which are in the single-lined spectroscopic binary sample. \citet{Gaia2023_FPR} compares the orbital periods from the FPR analysis to the \Gaia{} DR3 SB1 solution (their Figure 21) and find that the RV periods for the targets selected by Equation \ref{eqn:gaia_cut} agree within $10\%$ for all but one system. Seven of the remaining matches are in the single-lined second degree trend sample. Only one target in the \Gaia{} LPV FPR catalog, GDR3 2567779977831471232, has an astrometric orbit solution. Interestingly, the astrometric orbital period $P=927\pm85$~days for this system does not agree with the RV period in the \Gaia{} FPR catalog of $P=695$~days.

\citet{Gaia2023_FPR} show how the \nFPRtotal{} sources in the \Gaia{} FPR can be separated into pulsators and binaries by comparing the amplitude of the photometric variability, $A_G$, to the amplitude of the radial velocity variability, $A_{\rm{RV}}$, using the criteria that systems with
\begin{equation} \label{eqn:gaia_cut}
    A_G < 0.3\ \text{mag and} \ A_G < 0.25 \left(\frac{A_{RV}}{10\ \rm{ km\ s}^{-1}}\right)^2\ \text{mag}
\end{equation}

\noindent are binaries. Figure \ref{fig:target_selection} shows the distribution of $A_G$ and $A_{\rm{RV}}$ and the \nELLtotal{} targets selected as binaries using Equation \ref{eqn:gaia_cut}. The binary mass function
\begin{equation} \label{eqn:fm}
    f(M) = \frac{P K^3}{2\pi G} \left(1-e^2\right)^{3/2} = \frac{M_2^3 \sin^3 i}{(M_1+M_2)^2}
\end{equation}

\noindent is the absolute lower limit on the companion mass obtained in the limit of an edge-on inclination and a zero mass ($M_1=0\ M_\odot$) photometric primary. Figure \ref{fig:fm_dist} shows the distribution of mass functions for the binaries using the RV period and semi-amplitude from the \Gaia{} FPR catalog and assuming the eccentricity is $e=0$. There are \nHighFM{} binaries with $f(M) > 1.0\ M_\odot$ which are listed in Table \ref{tab:target_selection}.

\begin{figure*}
    \centering
    \includegraphics[width=\linewidth]{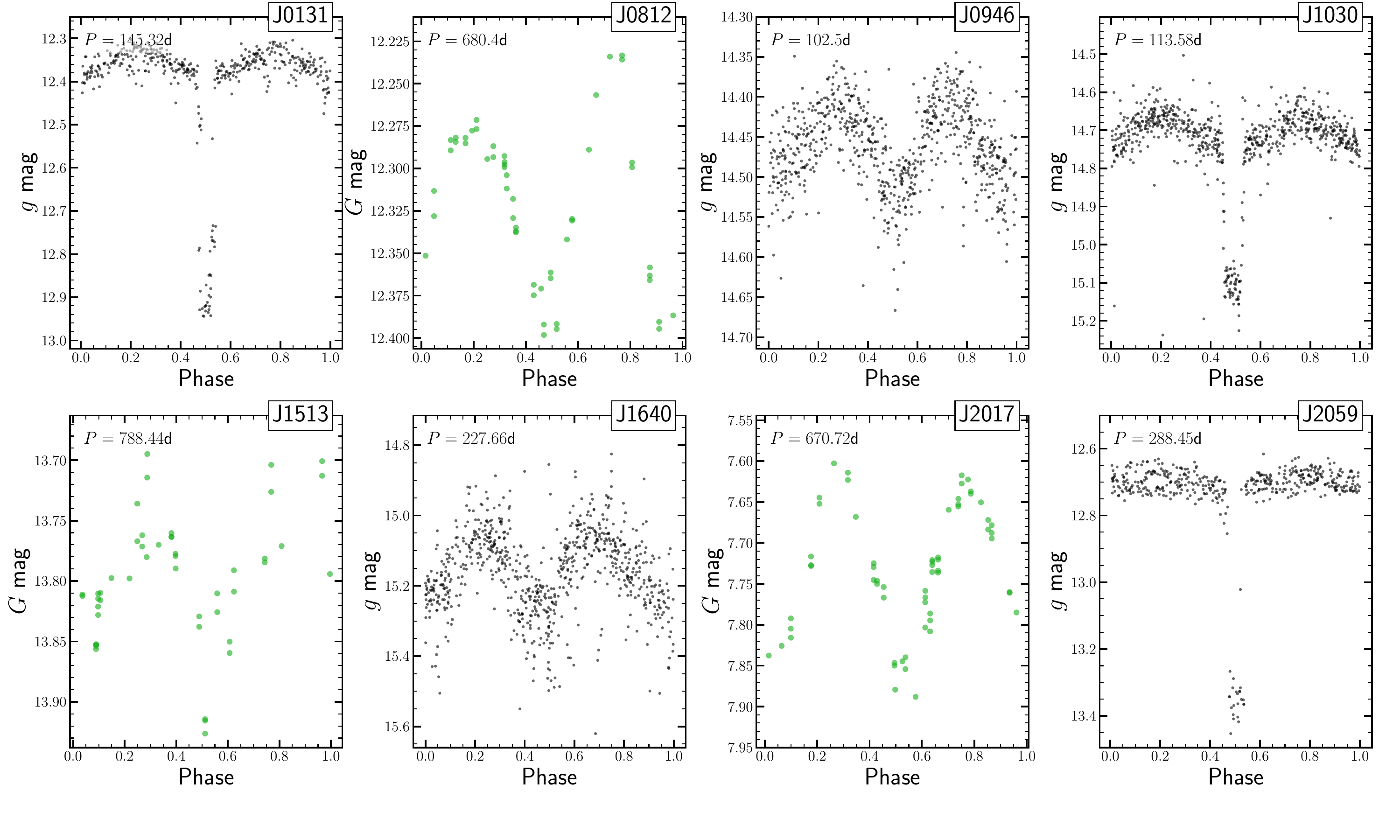}
    \caption{ASAS-SN $g$-band (black) and \Gaia{} $G$-band (green) light curves of the eight targets with $f(M)>1\ ~M_\odot$. Three targets (J0131, J1030, and J2059) show evidences of eclipses consistent with the spectroscopic period. The remaining targets are our candidate ellipsoidal variables.}
    \label{fig:lightcurves}
\end{figure*}

\subsection{Rejecting Eclipsing Binaries} \label{sec:eclipsingbinaries}

The targets in the \Gaia{} FPR catalog were selected based on their \Gaia{} photometric variability. The cadence and number of epochs in the \Gaia{} light curves are not well suited for identifying short, narrow eclipsing features in long-period binaries. To that end, we use light curves from the All-Sky Automated Survey for Supernovae \citep[ASAS-SN,][]{Shappee14, Kochanek17, Hart23} to identify eclipsing binaries in the sample of high mass function binaries. 

Out of the eight, targets, four are ASAS-SN variables (J0946, J2017, J1640, and J1030) and all were classified as semiregular variables \citep{Jayasinghe21, Christy23}. The ASAS-SN photometric period is also approximately half of the reported \Gaia{} RV period, as expected for ellipsoidal variables. Figure \ref{fig:lightcurves} shows the ASAS-SN light curves of six of the eight high $f(M)$ targets. We perform phase-dispersion minimization to refine the \Gaia{} period by searching a narrow period window ($0.05 P$) around the \Gaia{} period. 

Two targets, J0812 and J1513 do not have ASAS-SN Skypatrol V2 light curves. J0812 has a nearby ($<5\farcs0$) star, HD 68845, that is bright ($V\sim9$~mag) and blended with it in the ASAS-SN images. J1513 was omitted from the ASAS-SN Skypatrol V2 sample \citep{Hart23} because the ATLAS-REFCAT2 \citep{Tonry18} $g$-band magnitude is $g=17.9$~mag and the proximity statistic\footnote{The proximity statistic \texttt{r1} reports the radius where the cumulative flux from nearby stars exceeds the flux of the target.} is $<20\farcs0$. We recomputed the ASAS-SN $g$-band light curve using aperture photometry \citep{Kochanek17} and found that the target has a median $g=14.4$~mag. However, we find no evidence of periodic photometric variability using a Lomb-Scargle periodogram \citep{Lomb76, Scargle82}. J2017 is $g\sim10.2$~mag while ASAS-SN begins to saturate for stars brighter than $g\sim11$~mag. For these three targets we instead show the \Gaia{} light curves in Figure \ref{fig:lightcurves}. 

We remove J0131, J1030, and J2059 from consideration as non-interacting compact object binary candidates because of the presence of eclipses in their ASAS-SN light curves. The secondary eclipses in all three light curves are shallow or not visible, suggesting that the photometric secondary has a lower surface brightness. J0812, J0946, J1640, and J2017 have no evidence of eclipses and are likely ellipsoidal variables. J0812 also has uneven maxima in its \Gaia{} light curve, which could be evidence for the presence of a spot on the giant star. The \Gaia{} light curve of J1513 does have evidence for periodic photometric variability, but it is not a clear ellipsoidal variable. For all four targets, we find that the light curves phase with the \Gaia{} RVs as expected for binary stars.

We also inspect the \TESS{} light curves of our targets. As compared to the \Gaia{} and ASAS-SN light curves, the \TESS{} light curves are inherently less useful for characterizing these systems because of the 27-day \TESS{} Sector length. Many of the targets are also near the Galactic plane, so crowding presents an additional challenge in interpreting light curves given the large 21'' \TESS{} pixels. Even with these limitations, \TESS{} is still useful to search for low-amplitude features, such as shallow eclipses. We use three pipelines to retrieve \TESS{} light curves:
\begin{enumerate}
    \item The Quick-Look Pipeline \citep[QLP,][]{Huang20a, Huang20b}. These light curves are processed from the full-frame images (FFIs) and are available for targets brighter than $T=13.5$~mag. Since the detrending process can remove real variability on timescales $>0.3$~days, we use the raw, undetrended light curves.
    \item The Science Processing Operations Center \citep[SPOC,][]{Caldwell20}. These light curves are also generated from the FFIs, but are only available for 10,000 targets per Sector for each CCD with a magnitude limit of $T=13.5$~mag.
    \item \TESS{}-\Gaia{} Light Curve \citep[TGLC,][]{Han23}. This pipeline models the FFI with point-spread functions (PSFs) based on \Gaia{} DR3 positions. We use the calibrated PSF flux to reduce contamination from nearby variable targets.
\end{enumerate}

Since the long-period ELL or eclipsing binary (EB) signal produces a long-term trend in the \TESS{} light curves for each Sector, we do a simple check for prominent variable signatures by-eye to identify any shallow eclipsing features. J0812, J0946, J1513, J1640, J2017, and J2059 show no evidence of variability aside from the primary ELL/EB signal. The primary eclipse of J1030 is detected in the Sector 64 light curve with a phasing consistent with the ASAS-SN light curve. No secondary eclipse or other variable signals are observed. Finally, for J0131, we find some evidence of short period variability and additional eclipsing features outside of the $P=145$~day signal. We show this system's \textit{TESS} light curve and discuss why it is likely a blended target in Appendix~\ref{appendix:tess}.

\section{Radial Velocity Orbits} \label{sec:rvs}

\begin{table*}
    \centering
    \caption{RV orbit fit results for the five ELL candidates. The RV orbits are shown in Figure \ref{fig:rv_orbits}. The ratio of the orbital motion to the parallax is calculated using Equation \ref{eqn:orb_motion_parallax} assuming an edge-on inclination.}
    \sisetup{table-auto-round,
     group-digits=false}
    \renewcommand{\arraystretch}{1.5} % Adjust the value as needed
    \setlength{\tabcolsep}{12pt}
    \begin{center}
        \input{ANC/rvorbits_table}
    \end{center}
    \label{tab:rvorbits}
\end{table*}

The \Gaia{} FPR catalog reports the period and amplitude of the photometric variability but they do not fit a spectroscopic orbit model. Here, we fit the \Gaia{} time-series RVs using a prior based on the photometric period. We fit a Keplerian orbit model of the form
\begin{equation} \label{eqn:orbit}
    \text{RV}(t) = \gamma + K \left[(\omega+f)+e\cos\omega\right],
\end{equation}
\noindent where $\gamma$ is the center-of-mass velocity, $K$ is the radial velocity semiamplitude, $f$ is the true anomaly, and $\omega$ is the argument of periastron. The true anomaly, $f$, is related to the eccentric anomaly, $E$, and the eccentricity, $e$ by
\begin{equation}
    \cos f = \frac{\cos E - e}{1-e \cos E},
\end{equation}
and the eccentric anomaly is
\begin{equation}
    E - e\sin E = \frac{2\pi(t-t_0)}{P}
\end{equation}
where $P$ is the period and $t_0$ is the time of periastron. 

\begin{figure*}
    \centering
    \includegraphics[width=\linewidth]{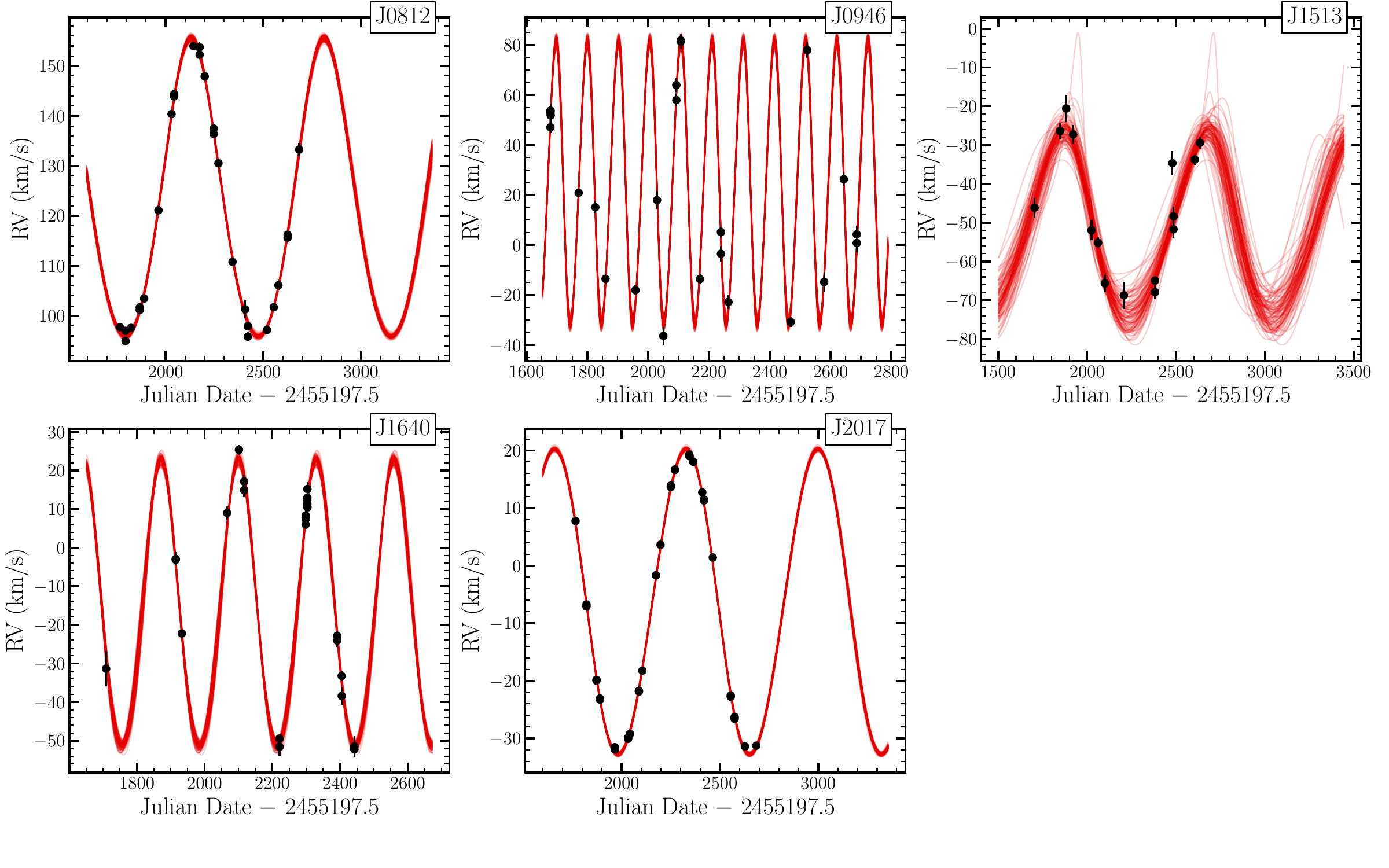}
    \caption{Radial velocity orbits for the five ELL variables with high $f(M)$ identified in the \Gaia{} FPR sample. The orbital parameters are reported in Table \ref{tab:rvorbits}.}
    \label{fig:rv_orbits}
\end{figure*}

We start by using \textsc{TheJoker} \citep{PriceWhelan17} for rejection sampling, using a Gaussian prior on the period based on the ELL period from the light curve. We use the resulting solution to initialize walkers for an MCMC fit with \textsc{emcee} \citep{Foreman-Mackey13}. Figure \ref{fig:rv_orbits} shows the RV orbits for the five ellipsoidal variable targets and Table \ref{tab:rvorbits} reports the orbital parameters. Two of the ELLs, J0812 and J0946, were also included in the \Gaia{} DR3 SB1 orbit catalog \citep{Gaia23_summary, Katz23}. For J0812, the \Gaia{} SB1 solution is $P_{\rm{SB1}} = 681\pm2$~days, $e_{\rm{SB1}}=0.077\pm0.008$, and $K_{\rm{SB1}}=29.9\pm0.2$~km/s, which agrees with our orbital solution (Table \ref{tab:rvorbits}). For J0946, the \Gaia{} SB1 solution is $P_{\rm{SB1}} = 102.60\pm0.07$~days, $e_{\rm{SB1}}=0.076\pm0.019$, and $K_{\rm{SB1}}=56.4\pm1.1$~km/s, which also agrees with our orbital fit. Although it is not surprising that we find the same orbital solution since we are fitting the same RV measurements, some of the \Gaia{} spectroscopic orbit solutions have been found to be spurious. \citet{Bashi22} use radial velocities from large surveys to predict a ``score'', $\mathcal{S}$, for each SB1 solution on a zero to one scale. J0812 and J0946 have $\mathcal{S}=0.92$ and $\mathcal{S}=0.71$, respectively, which are both well above the cutoff $\mathcal{S}>0.587$ that \citet{Bashi22} use for their ``clean'' sample.

The RV orbit of J1513 is less well-constrained. We searched for orbital solutions at different periods using rejection sampling with \textsc{TheJoker}, but were unable to find any other solution. \Gaia{} DR3 does not include an SB1 orbit for this target because the effective temperature of the RV template is 3500~K, below the cutoff of 3875~K \citep{Gaia23_hiddentreasure}. Since the light curve of this target is also the least consistent with ellipsoidal modulations, additional RV and photometric observations are needed to better characterize this system. We do not list this target as an ELL in Table \ref{tab:target_selection} and remove it from consideration as a non-interacting compact object candidate. 

\section{Properties of the Giant Stars} \label{sec:seds}

\begin{figure}
    \centering
    \includegraphics[width=\linewidth]{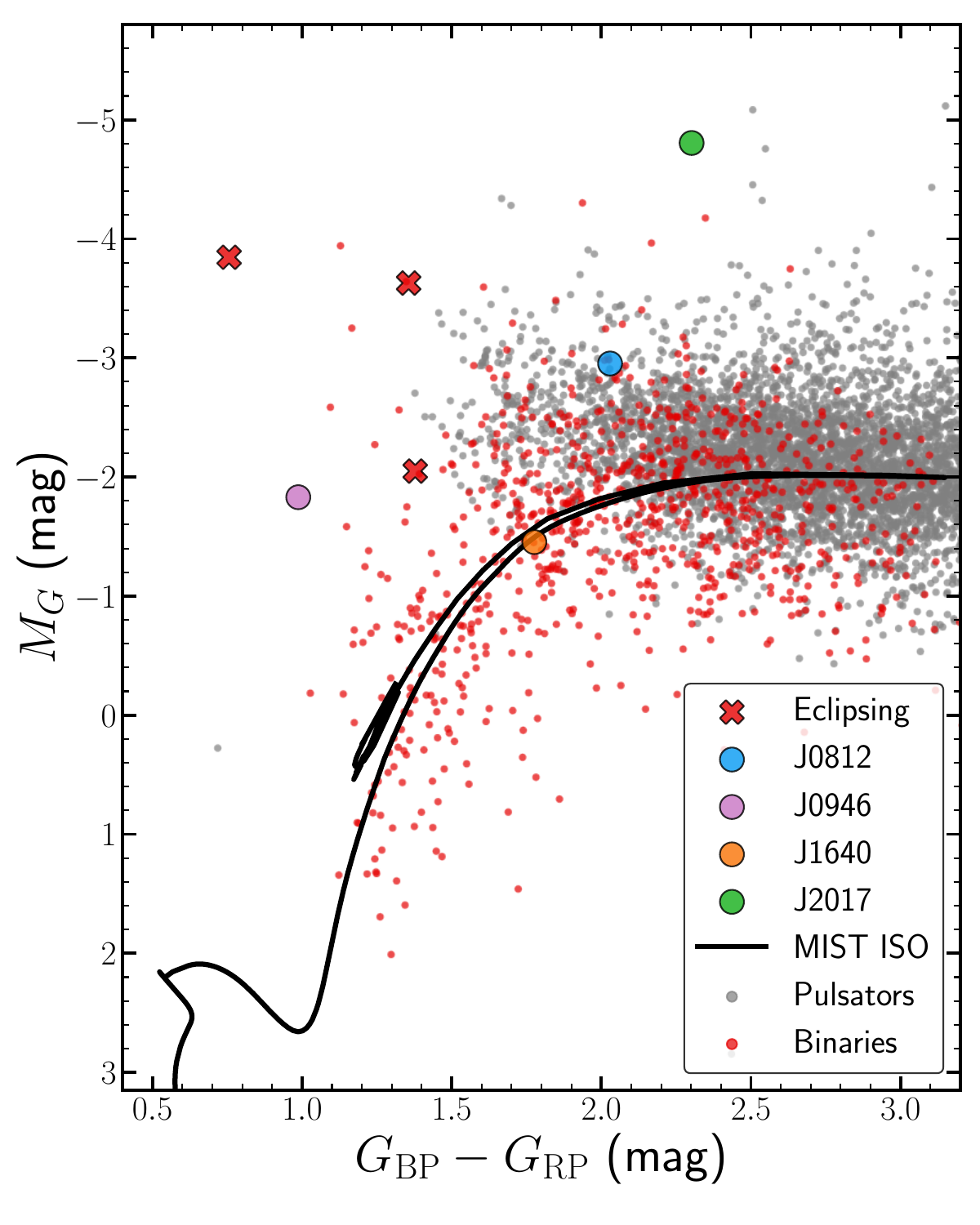}
    \caption{\Gaia{} color-magnitude diagram (CMD) for the \Gaia{} FPR catalog. Systems classified as pulsators (binaries) are shown in gray (red) based on the Equation \ref{eqn:gaia_cut}. The three high $f(M)$ eclipsing targets are marked as red crosses, and the four ELL targets are colored circles. J1513 is not shown because the \Gaia{} parallax uncertainty is $\varpi/\sigma_\varpi < 5$. The black line shows a MIST isochrone \citep{Dotter16, Choi16} for an age $10^{9.4}$~years.}
    \label{fig:cmd}
\end{figure}

To better understand the nature of the unseen companions we first need to characterize their photometric primaries. The \Gaia{} LPV catalog \citep{Lebzelter23}, the starting point for the \Gaia{} FPR catalog, includes a color filtering criteria of $G_{\rm{BP}}-G_{\rm{RP}} > 0.5$~mag. The color-magnitude diagram (CMD) in Figure \ref{fig:cmd} shows that all of the targets in the \Gaia{} FPR catalog are on the giant branch as expected. Absolute magnitudes are calculated using photogeometric distances from \citet{BailerJones21} and $V$-band extinctions from the {\tt mwdust} \citep{Bovy16} ``Combined19'' dust map \citep{Drimmel03, Marshall06,Green19}. We convert the $V$-band extinction to extinctions in the \Gaia{} filters using the coefficients from \citet{Wang19}. J1513 has a large fractional parallax error, $\varpi/\sigma_\varpi =2.3$, so its CMD position is uncertain and it is not shown in Figure \ref{fig:cmd}.

The \Gaia{} parallax measurements could also be biased by the orbital motion of the binaries for these systems. Following \citet{Thompson19}, we compare the orbital motion of the binary to the measured parallax as
\begin{equation} \label{eqn:orb_motion_parallax}
    \frac{\rm{orbital\ motion}}{\rm{parallax}} \approx \frac{0.42}{\sin i}\left(\frac{f(M)}{1.0\ M_\odot}\right)^{1/3}\left(\frac{P}{100\ \rm{days}}\right)^{2/3}.
\end{equation}
\noindent Table \ref{tab:rvorbits} includes this ratio for each target computed using the MCMC posteriors for the RV orbit assuming an edge-on inclination. For J0812, J1513, and J2017, the expected astrometric motion from the binary orbit is larger than the measured parallax. This could bias the stellar luminosity and radii in either direction depending on the projection of the binary on the sky. None of these targets have an astrometric orbit solution in \Gaia{} DR3. Only J2017 has a \Gaia{} renormalized unit weight error $\rm{RUWE} > 2$ (Table \ref{tab:target_selection}). 

The three eclipsing targets (J0131, J1030, and J2059) are bluer than the majority of the \Gaia{} FPR sample, which indicates that their companions are hotter subgiants or main sequence stars. The ELL J0946 also has extinction corrected color $G_{\rm{BP}}-G_{\rm{RP}}\approx 1.0$~mag, which suggests the presence of a luminous companion. 

Next, we fit the spectral energy distributions (SEDs) for three of the four remaining targets. We do not fit the SED of J0812 because it is blended with a nearby bright star. We use broad-band photometry from 2MASS \citep{Cutri03}, \WISE{} \citep{Wright10, Cutri12}, and the low-resolution \Gaia{} XP spectra \citep{DeAngeli23}. J1640 was observed by \textit{GALEX} \citep{Bianchi17} as part of the All-Sky Imaging Survey (AIS) in the near ultraviolet (NUV) band with an exposure time of 64 seconds. There is no source detected at the \Gaia{} position. We use {\tt gPhoton} \citep{Million16} to derive a 5$\sigma$ upper limit $\rm{NUV}< 20.6$~mag.

\begin{figure*}
    \centering
    \includegraphics[width=\linewidth]{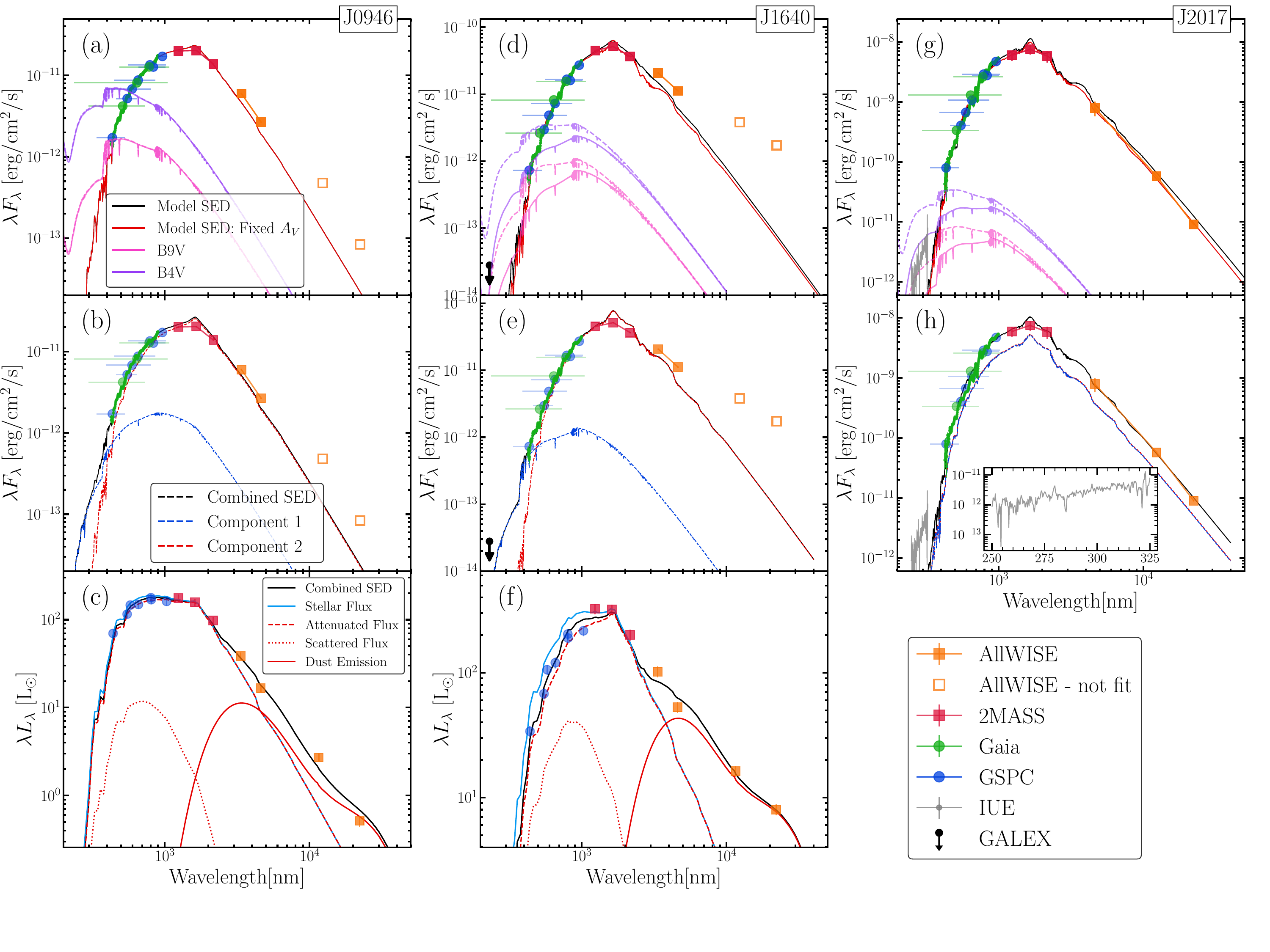}
    \caption{Spectral energy distributions (SEDs) of J0946, J1640, and J2017. The blue points in each panel show synthetic photometry from the \Gaia{} Synthetic Photometry Catalog \citep[GSPC,][]{Gaia23_GSPC}. For J1640, we show the upper limit for the \textit{GALEX} NUV band in panels (d) and (e). For J2017, we show the archival IUE spectrum in panels (g) and (h). The inset on panel (h) shows a zoomed-in view of the IUE spectrum. The top row shows the single-star SED fits. The red lines show the ``Fixed'' model where $A_V$ is fixed at the values from the 3-dimensional dust maps and the black lines show the ``Free'' model where extinction is a free parameter. For comparison, the purple and pink lines show the spectrum of a B4V and a B9V star, respectively extinguished by the same amount of dust as the free-$A_V$ model (solid lines) and fixed-$A_V$ model (dashed lines). The middle row (panels b, e, and h) show binary star SED fits using the fixed extinction from the 3-dimensional dust maps. The red and blue lines show the contribution from each component and the combined model is in black. J0946 and J1640 show evidence of significant IR excesses in the \textit{WISE} photometry. In panels (c) and (f) we show models including circumstellar dust computed using {\tt DUSTY} \citep{Ivezic99}. The combined model (black) is broken down into the intrinsic stellar flux (blue), the attenuated flux (dashed red), the scattered flux (dotted red), and dust emission (solid red). }
    \label{fig:sed_panel}
\end{figure*}

We fit the spectral energy distribution (SED) using the \citet{Castelli03} atmosphere models included in {\tt pystellibs}\footnote{\url{https://github.com/mfouesneau/pystellibs}}. We calculate synthetic photometry with {\tt pyphot}\footnote{\url{https://mfouesneau.github.io/pyphot/}} and sample over stellar parameters with {\tt emcee} \citep{Foreman-Mackey13} for 10000 iterations. We first fit the SED with the extinction from 3-dimensional dust maps ``Combined19'' dust map and use a Gaussian prior on the distance using the \citet{BailerJones21} photogeometric distance. Since J0946 and J1640 are in the southern sky, we use extinction maps from \citet{Lallement22} rather than the \citet{Drimmel03} map used by {\tt mwdust}. Next, we re-fit the SED with the extinction as a free parameter. We also fit for a binary star SED model with fixed extinction and distance. Finally, we consider a SED model with graphitic circumstellar dust using {\tt DUSTY} \citep{Ivezic97, Ivezic99}, following the procedure described in \citet{Neustadt23}, to model the IR excess for J0946 and J1640. Figure \ref{fig:sed_panel} shows the SED fits for the three targets and Tables \ref{tab:sed_table_j0946}--\ref{tab:sed_table_j2017} report the MCMC posteriors.

\begin{table}
    \centering
    \caption{SED fit results for J0946. We fit the SED with four models. The ``Free'' model includes extinction as a free parameter while the ``Fixed'' model uses the value from 3-dimensional dust maps. We also fit a binary model with two stars. The extinction and distance are fixed for this model. Finally, the ``Dust'' model includes circumstellar dust with optical depth $\tau_V$ and temperature $T_{\rm{dust}}$. $R_{\rm{dust}}$ is the distance from the giant to the dust shell. The core mass of the photometric primary is computed from Equation \ref{eqn:coremass} using the luminosity from the SED model and assuming a Solar metallicity. SED fits are shown in Figure \ref{fig:sed_panel}. We use the results of the model including circumstellar dust to set priors on the giant star in our \PHOEBE{} models (Section \S\ref{sec:phoebe}).}
    \sisetup{table-auto-round,
     group-digits=false}
    \renewcommand{\arraystretch}{1.5} % Adjust the value as needed
    \setlength{\tabcolsep}{6pt}
    \begin{center}
        \input{ANC/J0946/revised_sed_table_fullJ0946}
    \end{center}
    \label{tab:sed_table_j0946}
\end{table}

\begin{table}
    \centering
    \caption{Same as Table \ref{tab:sed_table_j0946}, but for J1640. We use the stellar parameters of the SED model with circumstellar dust to set priors on our \PHOEBE{} model in Section~\ref{sec:phoebe}.}
    \sisetup{table-auto-round,
     group-digits=false}
    \renewcommand{\arraystretch}{1.5} % Adjust the value as needed
    \setlength{\tabcolsep}{6pt}
    \begin{center}
        \input{ANC/J1640/revised_sed_table_fullJ1640}
    \end{center}
    \label{tab:sed_table_j1640}
\end{table}

\subsection{J0946: Gaia DR3 5405789050140488320}

\begin{figure*}
    \centering
    \includegraphics[width=\linewidth,trim={0 3cm 0 3cm},clip]{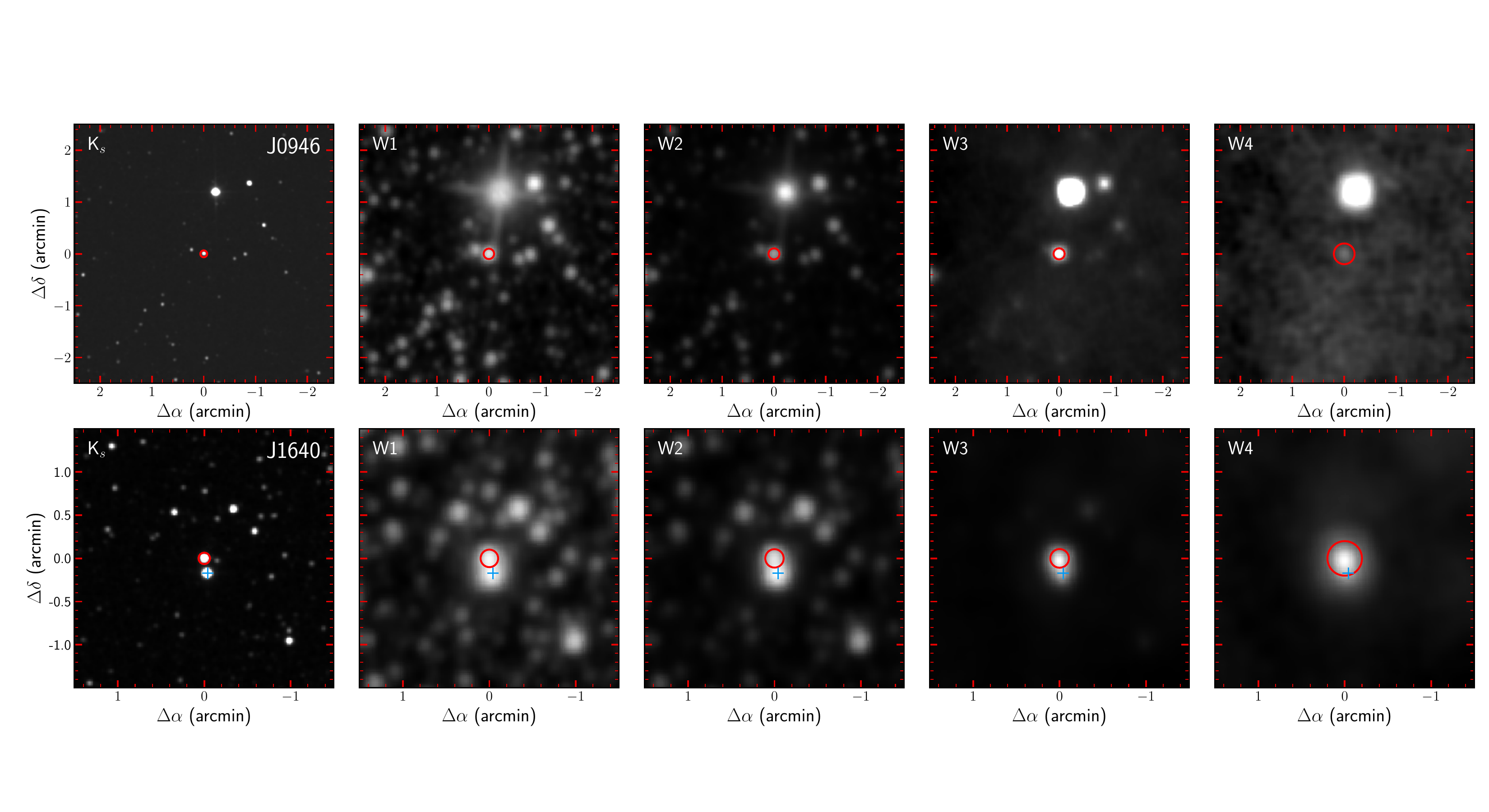}
    \caption{Cutouts of the 2MASS $K$-band and \textit{WISE} infrared images for J0946 (top) and J1640 (bottom). There is an IR bright star (IRAS 09450$-$5056, $\rm{W1}=5.4$~mag) $\approx 1\farcm2$ away from J0946. The diffraction spike may affect the photmetry of J0946 in the W1 and W2 bands. For J1640, there is a nearby star with apparent $\rm{W1}=7.4$~mag $\approx 10\farcs5$ away marked with the blue crosshairs. The circles in each panel show the approximate PSF for each filter \citep{Cutri03, Wright10}.}
    \label{fig:ir_cutouts}
\end{figure*}

The single-star SED fits for J0946 predict a stellar radius $R=33\pm1\ R_\odot$ and $T_{\rm{eff}}=4150^{+50}_{-40}$~K. The single star fits (Figure \ref{fig:sed_panel}a) show that a UV observation would rule out a massive main sequence companion of spectral type B4V ($M=5.1\ M_\odot$), but lower mass main sequence stars with $M\sim3\ M_\odot$ can still hide below the luminosity of the giant star. Both single-star SED fits are poor fits in the IR and near-IR, especially for the \textit{WISE} W3 and W4 filters, which we exclude when fitting the SED.

The two-star SED fit (Figure \ref{fig:sed_panel}b) does not improve the fit in the IR, but does show that the SED could be fit by a slightly cooler giant ($R_1=34.2^{+0.6}_{-0.7}\ R_\odot$, $T_{\rm{eff},1}=3980^{+30}_{-50}$~K) and a hotter, smaller secondary ($R_2=5^{+3}_{-2}\ R_\odot$, $T_{\rm{eff},1}=6200^{+800}_{-700}$~K). 

A fourth SED model including circumstellar dust is more consistent with the observed IR excess (Figure \ref{fig:sed_panel}c). In this model, the giant star is slightly cooler ($T_{\rm{eff}}=4447$~K) with a smaller radius ($R=28\ R_\odot$), as compared to the single-star fits. The interstellar extinction is lower for this model ($A_V=1.46$~mag), and consequently, the stellar luminosity is lower, $L=282\ L_\odot$. The circumstellar dust around J0946 could be produced in a radiatively driven wind which would dominate the observed spectrum in the IR for dust optical depths $\tau\gtrsim 0.1$ \citep{Ivezic95}.

First ascent late-type red giants can also have episodes of internal mixing that produce circumstellar dust associated with lithium enrichment \citep{deLaReza96, Jasniewicz99}. However, the timescale of these events is short ($\lesssim 10^5$~years), and only a few percent of giants are observed to be in this phase. Table \ref{tab:sed_table_j0946} reports the best fitting dust temperature, $T_{\rm{dust}} = 822$~K, and the optical depth at $\lambda=0.55\ \mu\rm{m}$ is $\tau_\nu = 0.18$. \citet{Mallick22} fit the SEDs of five Li-rich giants using a {\tt DUSTY} model with silicate grains. All five of their targets have average dust temperatures $<200$~K, much cooler than the best-fitting dust temperature for J0946. The optical depth in their systems is also smaller by a factor of 10. Hotter and more optically thick dust indicates that the dust is closer to the star and the mass loss episode occurred more recently or that the dust is forming in a wind. High-resolution spectra of J0946 could be used to measure the Li abundance and make more direct comparisons to the Li-rich giants with IR excesses. Some circumstellar dust could also be created during episodes of mass transfer \citep[e.g.,][]{Debes12, Lu13}. The large dust shell radius, $R_{\rm{dust}}=1400\pm300\ R_\odot$, indicates that the dust is actually circumbinary, providing further evidence for a history of mass transfer in addition to the circular orbits and ellipsoidal variability. 

It is also possible that the \WISE{} photometry has been contaminated by a nearby bright star. Figure \ref{fig:ir_cutouts} shows the 2MASS $K_s$ and \WISE{} W1--W4 image cutouts. IRAS~09450$-$5056 is a bright ($\rm{W1}=5.4$~mag) star $\approx 1\farcm2$ from J0946. The diffraction spike from IRAS 09450$-$5056 crosses J0946 in the W1 and W2 images. The W1 and W2 photometry measurements have an ``A'' quality flag, meaning that the source is detected with SNR>10, but J0946 is flagged with possible flux contamination from the diffraction spike in the W1--W3 bands. While the diffraction spike contamination is most relevant for variability analysis, and many targets are over-flagged \citep{Cutri13}, systematic errors in the \WISE{} photometry could explain the apparent IR excess.

If the mass transfer has stripped the photometric primary, the core mass represents an absolute lower limit on the primary mass. For giant stars with degenerate cores, the core mass $M_c$ is related to the luminosity $L$ by
\begin{equation} \label{eqn:coremass}
    \left(\frac{L}{\rm{L}_\odot}\right) \approx 1170\left(\frac{M_c}{0.4\ \rm{M}_\odot}\right)^{7}
\end{equation}
at Solar metallicity \citep{Boothroyd88}. Table \ref{tab:sed_table_j0946} includes the core mass for each SED model luminosity.

If we assume that the photometric primary is filling its Roche-lobe, $R_1 = R_{\rm{Roche}}$, we can use the Eggleton approximation for the volume-averaged Roche-lobe radius to find the minimum primary mass that would be consistent with the SED radius without Roche-lobe overflow. Starting from the \citet{Eggleton83} approximation for the Roche-lobe radius, we write the semimajor axis in terms of the Roche-lobe radius, $R_{\rm{Roche}}$, and the mass ratio, $q=M_2/M_1$, 
\begin{equation}
    a \approx R_{\rm{Roche}} \frac{0.6 q^{-2/3} + \ln(1+q^{-1/3})}{0.49 q^{-2/3}} = R_{\rm{Roche}}/E(q).
\end{equation}
The orbital period is also known, so 
\begin{equation}
    R_{\rm{Roche}}^3 = \frac{E(q)^3 G M_1(1+q) P^2}{4\pi^2}.
\end{equation}

\noindent This can be solved to determine the range of mass ratios where Roche-lobe overflow would be expected for a given primary mass. Figure \ref{fig:egg} shows this radius relative to the radius estimates $R_{\rm{SED}}$ for the dusty SED model of J0946 and for the SED models of J1640 and J2017. For the J0946 system, we find that $R_{\rm{SED}} < R_{\rm{Roche}}$ for all mass ratios, which means that the giant can be entirely stripped without overflowing its Roche-lobe.

\subsection{J1640: Gaia DR3 5968984160290449024}

The SED of J1640 also shows a strong IR excess (Figure \ref{fig:sed_panel}d). As with J0946, we exclude the W3 and W4 filters from the single-star and binary-star SED fits that do not include circumstellar dust (Panels d and e of Figure \ref{fig:sed_panel}). As compared to J0946, J1640 is cooler and larger, though the estimated parameters differ significantly between SED models (Table \ref{tab:sed_table_j1640}). For the model where the extinction is fixed from the 3-dimensional dust maps, the giant is predicted to have radius $R=59^{+4}_{-5}\ R_\odot$ and effective temperature $T_{\rm{eff}}=3630^{+20}_{-20}$~K. The model with free extinction prefers substantially more interstellar extinction, $A_V=3.39$~mag as compared to $A_V=2.44$~mag, and has a larger stellar luminosity and higher temperature as a result. 

Figure \ref{fig:egg} shows the Roche-lobe radius relative to the SED radius for the fixed extinction model and the model with circumstellar dust. For the fixed extinction model, if the primary is entirely stripped, the stellar radius is larger than the maximum radius before Roche-lobe overflow for $q<1$. The model including circumstellar dust predicts a smaller stellar radius, and the giant can be totally stripped without overflowing its Roche lobe for this model.

Even with a \textit{GALEX} upper limit, the high extinction limits our ability to rule out massive main sequence companions. Figure \ref{fig:sed_panel}d shows the SED of a B4V and B9V with the same amount of extinction found in the SED models. While we can rule out a B4V companion in the model with fixed extinction, a B4V companion could not be ruled out if the extinction is $A_V=3.39$~mag. For either model, we find that a B9V star would be consistent with the upper limit. 

The binary star solution for J1640 suggests a hotter, less luminous companion with a luminosity ratio $<0.1$. A \textit{Swift} UVM2 observation (effective wavelength $\lambda_{\rm{eff}}=225$~nm) could be used to place stronger UV limits and search for luminous main sequence/subgiant companions, though this is made challenging due to the high extinction towards this target (see Section \S\ref{sec:discussion}).

As with J0946, only the model including circumstellar dust is able to describe the IR emission of J1640 (Figure \ref{fig:sed_panel}f). The dust temperature is $T_{\rm{dust}}=630^{+30}_{-20}$~K, and the optical depth is $\tau_\nu=0.68^{+0.09}_{-0.08}$, suggesting a moderate amount of dust absorption and scattering. The dust in this system is also circumbinary with $R_{\rm{dust}} = 4000\pm500\ R_\odot$. We use the stellar parameters from this model to set priors on our light curve fit in Section \S\ref{sec:phoebe}.

There is also evidence for systematic issues with the \WISE{} photometry of J1640. Figure \ref{fig:ir_cutouts} shows the IR images of J1640. There is a second bright \WISE{} target, ALLWISE J164023.34$-$411619.9 $\approx 10\farcs5$ away from J1640. In the 2MASS $K$-band image the two stars are clearly resolved but they are blended in the \WISE{} W1--W4 photometry. The \WISE{} catalog flags the W1 and W2 photometry as being potentially contaminated by this star. The star is also a long-period photometric variable in \Gaia{} DR3 \citep[GDR3 5968984164630392960,][]{Lebzelter23} with a photometric period of $P=753$~days, but it was not included in the \Gaia{} Focused Product release sample. While the two stars may be blended in the \WISE{} photometry, the separation is large enough that no contamination is expected in the ASAS-SN light curves or the \Gaia{} radial velocities. 

\begin{table}
    \centering
    \caption{Same as Table \ref{tab:sed_table_j0946}, but for J2017. Since this target does not have an IR excess we do not fit a model including circumstellar dust. Since this target has luminosity $>10^4\ L_\odot$, we compute the core mass using Equation \ref{eqn:core_mass_rsg} \citep{Paczynski70} for red supergiant stars.}
    \sisetup{table-auto-round,
     group-digits=false}
    \renewcommand{\arraystretch}{1.5} % Adjust the value as needed
    \setlength{\tabcolsep}{10pt}
    \begin{center}
        \input{ANC/J2017/sed_table_fullJ2017}
    \end{center}
    \label{tab:sed_table_j2017}
\end{table}

\subsection{J2017: Gaia DR3 2053893497434322048}

\begin{figure}
    \centering
    \includegraphics[width=\linewidth]{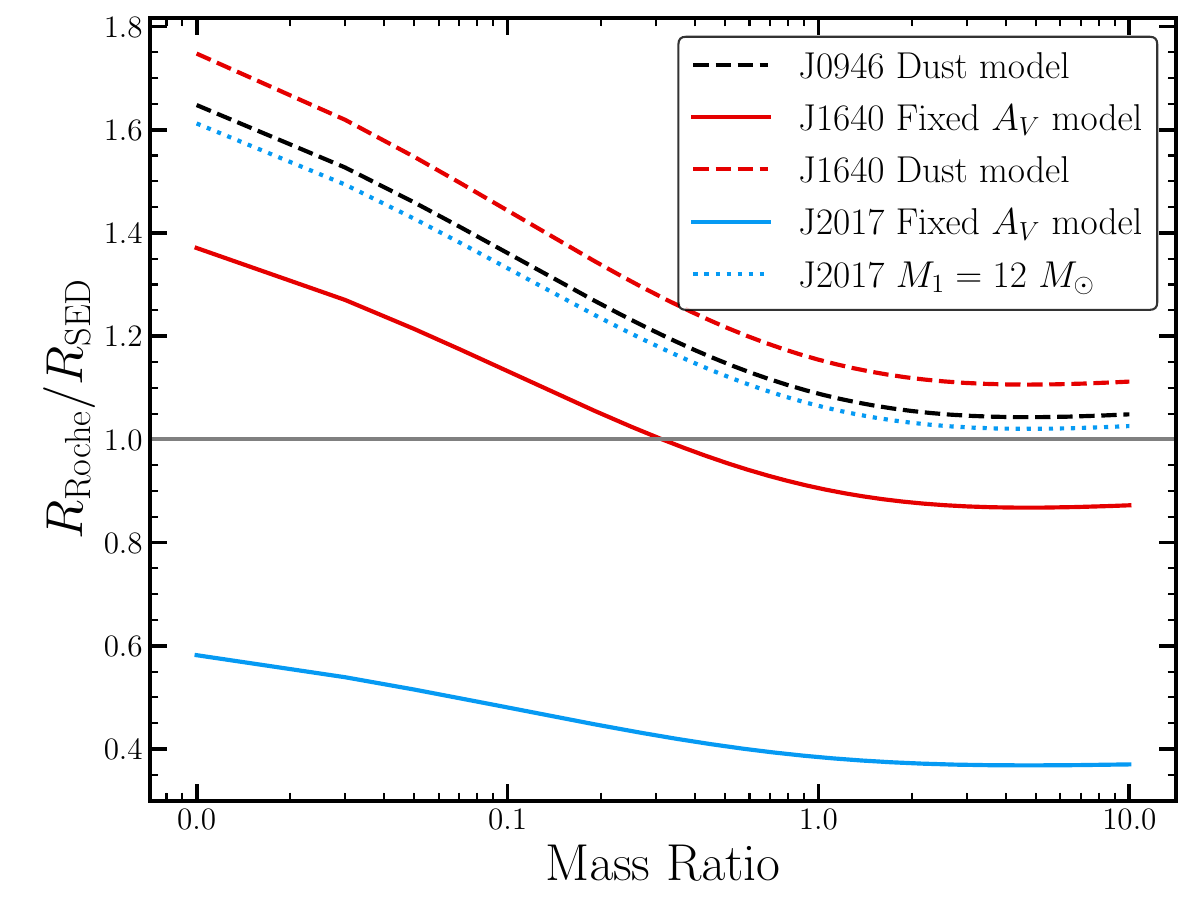}
    \caption{Roche-lobe radius relative to the radius from the SED, $R_{\rm{SED}}$, as a function of mass ratio, $q$. Where the curve is above 1, the primary can have the given mass and SED radius without Roche-lobe overflow. Systems below the gray line have giants that overflow their Roche-lobes.}
    \label{fig:egg}
\end{figure}

J2017 (V2012 Cyg, HD 332077) is one of the most luminous targets in the \Gaia{} FPR LPV catalog (Figure \ref{fig:cmd}), and the single-star SED models both suggest that the photometric primary has luminosity $>10^4\ L_\odot$ (Figure \ref{fig:sed_panel}g). The model where extinction is a free parameter predicts a higher extinction than the value from the {\tt mwdust} extinction maps, and consequently predicts a higher stellar luminosity and larger radius. The binary star SED model for J2017 prefers a solution with a luminosity ratio near unity (Figure \ref{fig:sed_panel}h). This model can be rejected because J2017 is observed as a single-lined spectroscopic binary. Unlike J0946 and J1640, J2017 has no IR excess, so we do not fit a model that includes circumstellar dust.

J2017 has previously been characterized as an `S-star' \citep{Stephenson84}, a transition object between M giants and carbon stars. Many S-stars show evidence of technetium (Tc) in their atmospheres, which is produced in the s-process and is short lived. However, no Tc is observed in the atmosphere of J2017 \citep{Jorissen92}, and Tc-deficient S-stars have been shown to have a higher binary fraction \citep{Brown90}.

Based on the SED fits, the luminosity of J2017 is $>10^4\ L_\odot$ while equation \ref{eqn:coremass} is only valid for lower mass stars with luminosities $\lesssim 2000\ L_\odot$. For red supergiants with carbon and oxygen cores, the core mass-luminosity relation is
\begin{equation} \label{eqn:core_mass_rsg}
    \left(\frac{L}{\rm{L}_\odot}\right)\approx 59250\left(\frac{M_c}{\rm{M}_\odot} - 0.522 \right)
\end{equation}
\noindent \citep{Paczynski70}. Hot bottom burning provides extra luminosity, breaking the linear relationship \citep[e.g.,][]{Wagenbuber98}, but we use the \citet{Paczynski70} relation as a first order estimate. Table \ref{tab:sed_table_j2017} reports the core mass of J2017 using this relation.

The giant primary must be fairly massive in order to have a luminosity $>10^4\ L_\odot$. A $10\ M_\odot$ star with Solar metallicity spends $\sim 2.8$~Myr with $L > 10^4\ L_\odot$ based on MIST evolutionary tracks \citep{Choi16, Dotter16}. For comparison, a $5\ M_\odot$ star has $L>10^4\ L_\odot$ for only $\sim 56,000$~years. The large SED radius, $R_{\rm{SED}} = 330^{+30}_{-30}\ R_\odot$ for the fixed {\tt mwdust} extinction model, also requires a large primary mass. Figure \ref{fig:egg} shows that the ratio $R_{\rm{Roche}} / R_{\rm{SED}} < 1$ unless the photometric primary is $\gtrsim 12\ M_\odot$. 

\citet{Jorissen92} use the spectroscopic orbit and an upper limit from the International Ultraviolet Explorer \citep[IUE,][]{Boggess78} to suggest that the companion to J2017 is an A-star. Follow-up low resolution spectra taken with the International Ultraviolet Explorer \citep[IUE,][]{Boggess78} found evidence of an A-type star spectra from 173--198~nm, but \citet{Ake92} note that a simultaneous fit to the long and short wavelength regions with a A-star plus giant star model is not successful. We include the IUE spectrum of the long wavelength region from May 1996 for comparison in Figure \ref{fig:sed_panel}g and a zoomed-in view in the inset of Figure \ref{fig:sed_panel}h. The flux from the IUE spectrum exceeds what is expected from our SED model, suggesting that there is a luminous companion. There is also evidence of MgII line emission at $\lambda280$~nm, which is associated with chromospheric activity in FGK stars \citep{Buccino08}.

A main sequence A6 companion with mass $M_2=1.8\ M_\odot$ would not produce the observed radial velocity semi-amplitude if $M_1> 12\ M_\odot$, as suggested by our SED fits. \citet{Jorissen92} instead suggest that the giant is very low mass $<0.5\ M_\odot$ with $R_1 < 245\ R_\odot$. This radius is inconsistent with our SED results (Table \ref{tab:sed_table_j2017}). 

Our picture of a massive giant primary can be reconciled with the stripped star+A-star scenario proposed by \citet{Jorissen92} if the \Gaia{} parallax is biased by the binary motion. The \Gaia{} parallax is $\varpi=0.4930\pm0.0535$~mas after using the zero-point correction from \citet{Lindegren21}. The astrometric signal from the binary motion is then $\simeq 0.801\ \rm{mas}/sin(i)$ (Equation \ref{eqn:orb_motion_parallax}). While a more detailed numerical simulation of the orbit incorporating the \Gaia{} scanning law and observation times is needed to fully understand the parallax bias \citep[e.g.,][]{Thompson19}, if the true \Gaia{} parallax is larger than the reported value, the actual stellar radius will be smaller than the result from our SED fit. 

New ultraviolet observations could be used to search for additional evidence of a luminous companion and validate the A-star signal identified by \citet{Ake92}. Our SED fits suggest that a main sequence A6V star would exceed the flux from the giant in the \textit{Swift} UVM2 filter. However, the A-star companion would have a magnitude of $\sim 20.1$~mag in the UVM2 band using the extinction and distance from the fixed extinction model (Table \ref{tab:sed_table_j2017}), which is approaching the sensitivity limit for the instrument \citep[e.g.,][]{Molina20} unless the binary motions have biased the parallax. While we are unable to provide a definitive characterization of this system here, it is unlikely that the companion is a compact object. We remove this target from further consideration.

\section{Models of Ellipsoidal Variability} \label{sec:phoebe}

\begin{figure*}
    \centering
    %\captionsetup{justification=centering}
    \includegraphics[width=0.9\linewidth]{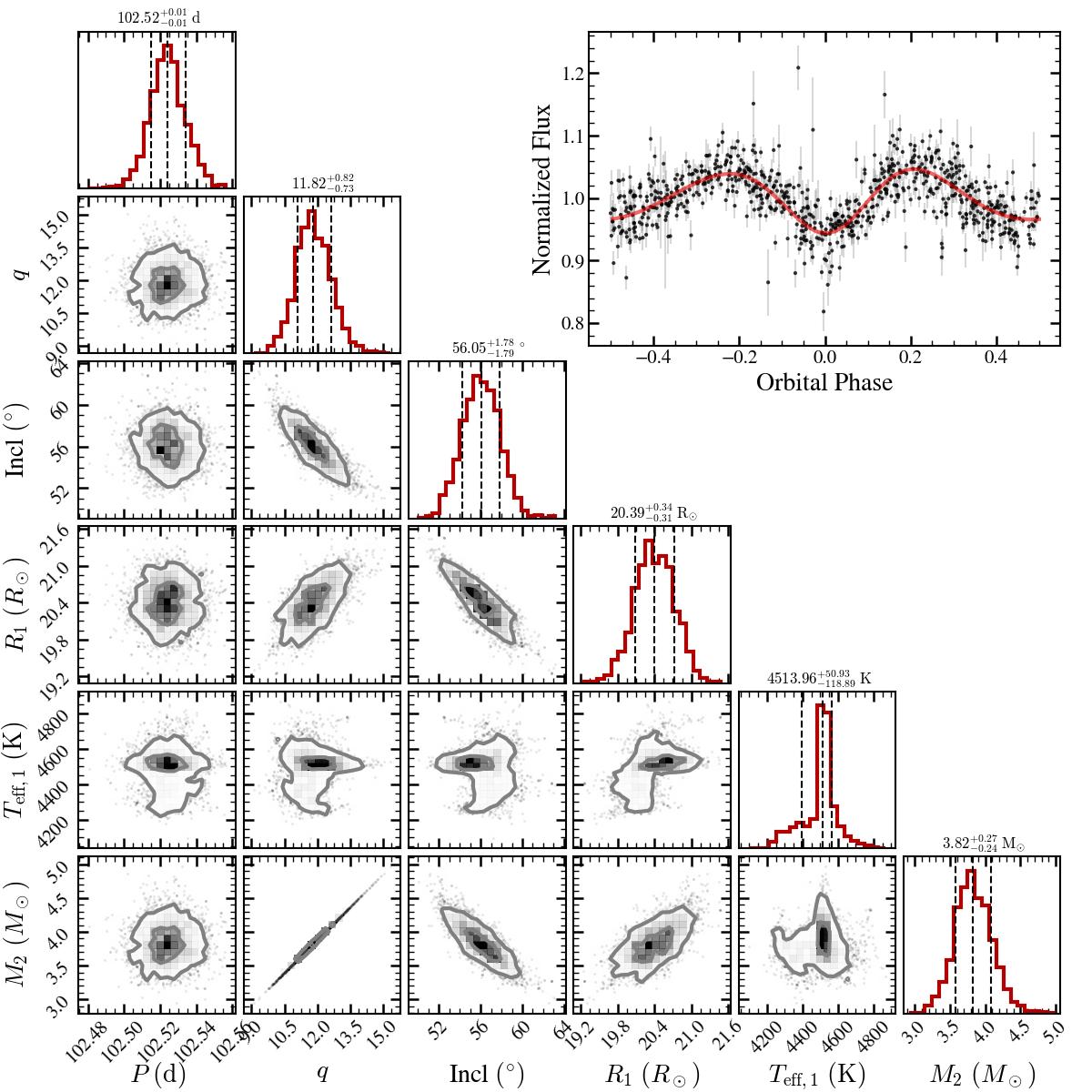}
    \caption{MCMC posteriors and light curve fit for J0946.}
    \label{fig:J0946_corner}
\end{figure*}

\begin{figure*}
    \centering
    %\captionsetup{justification=centering}
    \includegraphics[width=0.9\linewidth]{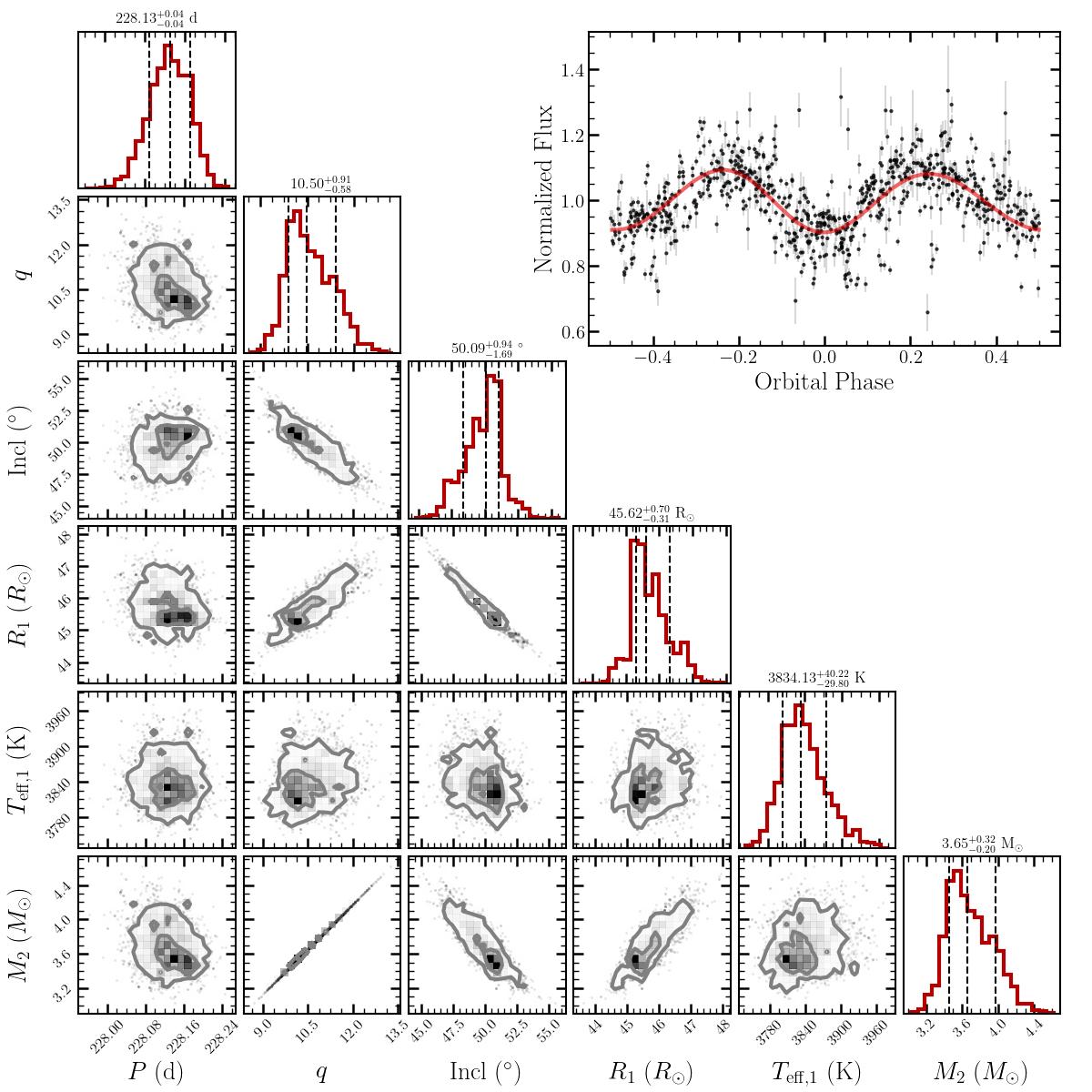}
    \caption{MCMC posteriors and light curve for J1640.}
    \label{fig:J1640_corner}
\end{figure*}

By modeling the ellipsoidal variable light curve and radial velocity curve simultaneously, we can constrain the orbital inclination and mass of the unseen companion. We used PHysics Of Eclipsing BinariEs Eclipsing Binary \citep[\PHOEBE{},][]{Prsa05, Conroy20} to fit the ASAS-SN light curves and \Gaia{} radial velocities for J0946 and J1640. We exclude J2017 because of the evidence of an A-star companion in the UV spectrum \citep{Ake92}.

For J0946 and J1640 we start by using \PHOEBE{} to fit a ``dark companion'' model, where the secondary is fixed to be small $R_2=3\times 10^{-6}\ R_\odot$, and cold, $T_{\rm{eff,2}}=300$~K, and ignore effects of irradiation and reflection. We fix the mass of the photometric primary to be the core mass $M_c$ computed using the SED model with circumstellar dust (Tables \ref{tab:sed_table_j0946} and \ref{tab:sed_table_j1640} and Equation \ref{eqn:coremass}) so that the resulting companion mass is the lower limit for a maximally stripped red giant.

To obtain an initial estimate for the fit, we use the \PHOEBE{} differential evolution optimizer. For the MCMC fit, we set Gaussian priors on the effective temperature and photometric primary radius from the SED fit and a prior on the period from the \Gaia{} FPR RV solution. We run the \PHOEBE{} models for 5000 iterations and use 26 walkers. We visually inspect the walker distributions to select a burn-in period for each target.  

We also run a second set of \PHOEBE{} models where the companion is luminous and the effective temperature $T_{\rm{eff, 2}}$ and radius $R_2$ are allowed to vary. The photometric primary mass $M_1$ is also a free parameter in this model and 32 walkers are used for the MCMC fit. 

\subsection{J0946: Gaia DR3 5405789050140488320}

Figure \ref{fig:J0946_corner} shows the corner plot for the dark companion \PHOEBE{} model of J0946. Table \ref{tab:phoebetable} reports the MCMC posteriors. As compared the circumstellar dust SED model, the dark companion \PHOEBE{} model predicts a smaller giant ($R=20.4^{+0.3}_{-0.3}\ R_\odot$) with an effective temperature that agrees with this SED within $1\sigma$. The mass ratio is large $q>10$, and the companion has $M_2=3.8^{+0.3}_{-0.2}\ M_\odot$. Based on the SEDs (Figure \ref{fig:sed_panel}), ultraviolet observations could be used to rule out main sequence stars of this mass.

The \PHOEBE{} model where the companion is treated as a luminous star finds a solution with two red giants. The orbital elements and masses are similar between the two models. The photometric primary in this model has $R_1=28^{+1}_{-1}\ R_\odot$ and $T_{\rm{eff,1}}=4430^{+60}_{-150}$~K, which agrees with the circumstellar dust SED model. The companion is also large $R_2=22^{+3}_{-2}\ R_\odot$ and hotter than the primary. The luminosity ratio this system is $L_2/L_1=0.85$. While it seems likely that a companion with this luminosity ratio would be detected in the \Gaia{} RVS spectra, the spectral lines of the smaller hotter companion could be broad and not detected if it is rapidly rotating. For comparison, V723 Mon, has a $L_2/L_1=0.74$ and is only detected as a SB1 since the projected rotational velocity of the companion is high \citep[$v_2\sin i=70\pm10$~km/s,][]{Jayasinghe21_unicorn, ElBadry22_zoo}. If the companion is also a luminous giant with this radius and effective temperature, its flux would exceed that of the photometric primary at wavelengths shorter than $\sim 6000$~\AA, so additional optical spectra could be used to search for evidence of rotationally broadened spectral lines.

\subsection{J1640: Gaia DR3 5968984160290449024}

Figure \ref{fig:J1640_corner} shows the corner plot and light curve fits for the dark companion \PHOEBE{} models of J1640. For both the dark and luminous companions models, the radius of the giant primary agrees with the results from the circumstellar dust SED with within $1\sigma$. The dark companion \PHOEBE{} model predicts an effective temperature of the giant $T_{\rm{eff,1}}=3830^{+40}_{-30}$~K, which is within $2\sigma$ of the SED result. The effective temperature of the giant in the luminous companion model agrees with the SED result at the $1\sigma$ level.

The orbital inclination in both of the \PHOEBE{} models for J1640 is predicted to be $\sim 50^{\circ}$ and the companion mass $M_2=3.7\ M_\odot$. As with J0946, if the unseen companion is actually a main sequence star with $M_2=3.7\ M_\odot$ it could be searched for with ultraviolet observations, though this would be more challenging for J1640 due to the high extinction ($A_V=2.44$~mag from the \citet{Lallement22} maps). In the luminous companion model, the secondary is a cool $T_{\rm{eff,2}}=3800^{+400}_{-200}$, smaller star $R_2=4^{+4}_{-2}\ R_\odot$, but the secondary radius is poorly constrained. The luminosity ratio is small, $L_2 / L_1 = 0.006$ but we note that the properties of such a star are not consistent with expectations from single star evolution. A star with $M_2=3.7\ M_\odot$ would cool to $3700$~K after it evolves off the main sequence, at which point its radius would be $\gtrsim 10\ R_\odot$. 

%For J1640, the radius is consistent between the \PHOEBE{} model and the SED model with circumstellar dust, but the effective temperature is lower for the \PHOEBE{} model ($T_{\rm{eff}}=3830^{+40}_{-30}$~K) than the result from the SED model. Both targets predict companion masses $\sim 3.7\ M_\odot$ and moderate inclinations $\sim 50^{\circ}$. Based on the SEDs (Figure \ref{fig:sed_panel}), ultraviolet observations of these targets could be used to rule out main sequence stars of this mass. However, subgiant or lower giant stars that have accreted mass from the photometric primary could conceivably still ``hide'' in the SEDs of the giant stars. 

\begin{figure}
    \centering
    \includegraphics[width=\linewidth]{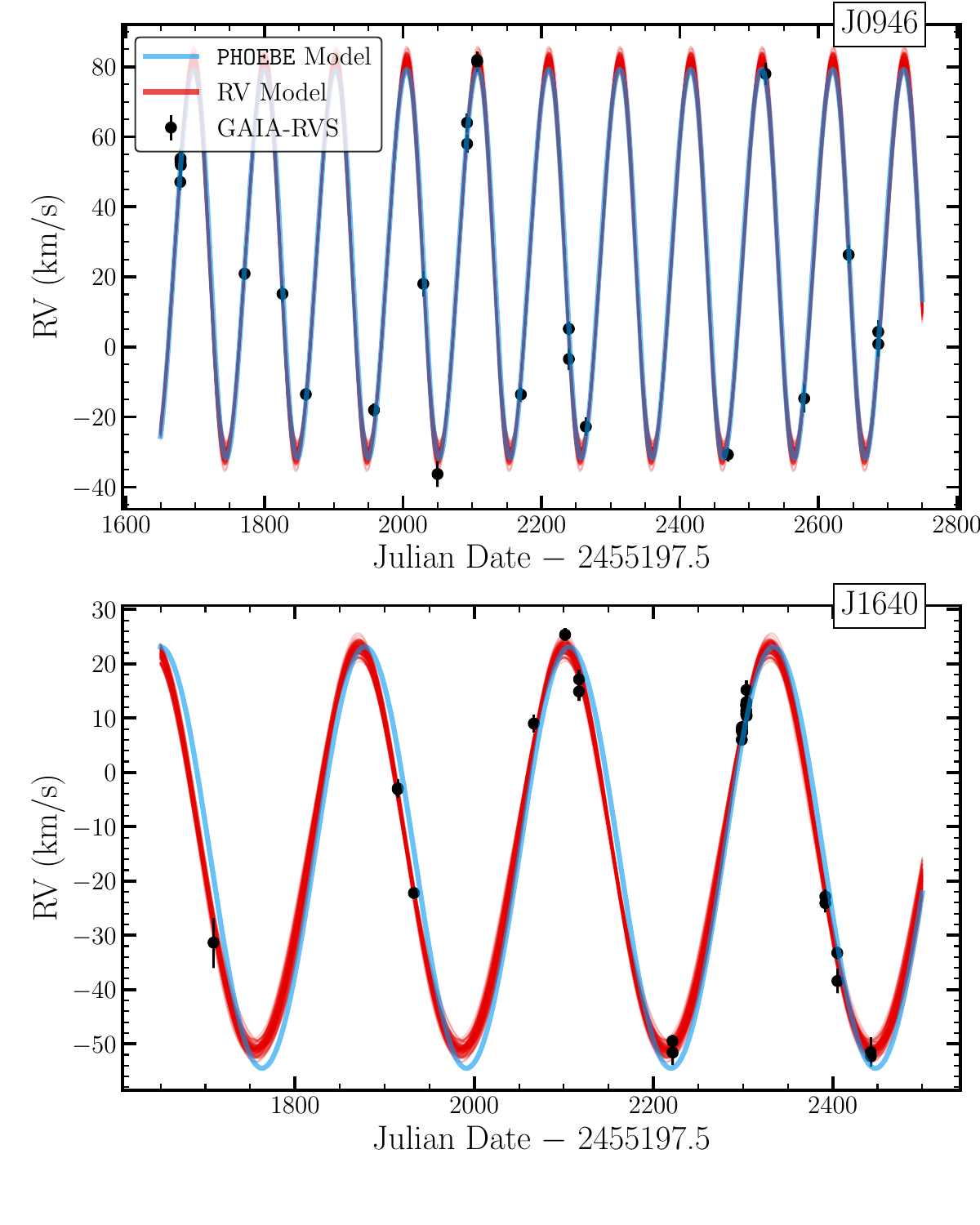}
    \caption{Comparison of the RV orbit fit from the Keplerian model (Section~\S\ref{sec:rvs}) and the joint \PHOEBE{} model fit to the ASAS-SN photometry and \Gaia{} RVs. While the two models produce a similar fit for J0946 (top), there is a significant offset between the two models of J1640 (bottom). This could be due to apsidal motion or a subset of poor RV measurements (Section \S\ref{sec:phoebe}).}
    \label{fig:rvmodel_comparison}
\end{figure}

The \PHOEBE{} models of J1640 predict near circular orbit $e=0.01$ ($e=0.03$ for the luminous companion model), which is significantly smaller than the ellipticity in our RV orbit model from Section \ref{sec:rvs} ($e=0.05$). Figure \ref{fig:rvmodel_comparison} shows the RV orbit model from Section \S\ref{sec:rvs} compared to the \PHOEBE{} result. While the ASAS-SN photometry is well-fit by the \PHOEBE{} model (Figure \ref{fig:J1640_corner}), the \PHOEBE{} RV model does not fit the \Gaia{} RVs. This suggests that there is a small relative phase offset between the RV orbit and the ellipsoidal modulations. 

Apsidal motion could produce a relative phase offset between the \Gaia{} RVs and the ellispodial modulations. Since the \PHOEBE{} model prefers a circular eccentricity, the argument of periastron, $\omega$, is unconstrained in the \PHOEBE{} model of J1640. If we instead adjust $\omega$ of the RV model from Section \S\ref{sec:rvs} to match the phasing of the \PHOEBE{} model, we find a change in argument of periastron $\Delta\omega\sim15^{\circ}$ is needed. The median time of the ASAS-SN $g$-band light curve is $\approx5.2$~years later than the median time of the \Gaia{} RVs, so $\Delta\omega / \Delta t\simeq 3\ \rm{deg\ year}^{-1}$. For a synchronized binary, \citet{Sterne39} and \citet{Rosu20} write the rate of apsidal motion as,

\begin{alignat}{2}
\dot\omega = \frac{2\pi}{P} \Biggl[&15f(e) \biggl(&& k_1 q \Bigl(\frac{R_1}{a}\Bigr)^5 + \frac{k_2}{q}\Bigl(\frac{R_2}{a}\Bigr)^5 \biggr) \notag\\
&+g(e) \biggl(&& k_1 q \Bigl(1 + \frac{1}{q}\Bigr) \Bigl(\frac{R_1}{a}\Bigr)^5 +\notag\\
& && k_2 \Bigl(1+\frac{1}{q}\Bigr) \Bigl(\frac{R_2}{a}\Bigr)^5\biggr)\Biggr],\notag \\
\end{alignat}
\noindent where $k_1$ and $k_2$ are the second order apsidal motion constants for the primary and the secondary, respectively, and
\begin{equation}
    \begin{split}
        f(e) &= \frac{ 1+ \frac{3e^2}{2} + \frac{e^4}{8}}{(1-e^2)^5},\ \rm{and} \\
        g(e) &= \frac{1}{(1-e^2)^2}.
    \end{split}
\end{equation}
There is also a contribution to $\dot{\omega}$ from general relativity, but for this system $\dot{\omega}_{\rm{GR}} \ll \dot{\omega}$ and it can be neglected. The values of $k_1$ and $k_2$ depend on the interior structure of the stars. If the companion is a main sequence star or a compact object we effectively have $k_2=0$. Using the period and eccentricity from the RV fit (Table \ref{tab:rvorbits}), the values of $R_1$ and $M_1=M_c$ from the SED (Table \ref{tab:sed_table_j1640}), and setting the companion to be a $3.7\ M_\odot$ BH, the value of $k_1$ would need to be $k_1\simeq0.145$ in order to have $\dot{\omega} \simeq 3\ \rm{deg\ year}^{-1}$. This is roughly twice as large as the value of $k_1$ estimated for the stripped star 2M04123153+6738486 \citep{ElBadry22_zoo}, and is outside of the range of the theoretical grids from \citet{Claret19}. While apsidal motion could produce some change in $\omega$ in the time between the \Gaia{} RVs and the ASAS-SN photometry, it is unlikely to be responsible for the observed phase offset. If J1640 is actually a hierarchical triple, light travel time effects (LTTE) could contribute to the effects of apsidal motion and increase the apparent phase offset between the RVs and the ellipsoidal variations. There are triple systems with apsidal motion \citep[e.g.,][]{Bozkurt07}, but these are generally short-period ($P<10$~day) inner binaries. 

It could also be the case that one or more \Gaia{} RVs are poorly constrained or have underestimated uncertainties. We test this by re-fitting the RV model from Section \S\ref{sec:rvs} with {\tt emcee} dropping all possible combinations of one, two, or three RV measurements. The initial walker positions for all MCMC runs are set using the rejection sampling results from {\tt TheJoker} using the full set of RVs.

For each trial, we phase the ASAS-SN light curve using the period and the time of superior conjunction. To determine the deviation from the expected phasing, we fit an analytic model of the form
\begin{equation}
    F(\phi) = a\cos(\phi) + b\cos(2\phi) + c\cos(3\phi) + d
\end{equation}
\noindent motivated by the analytic model of \citet{Morris93} and compute the reduced $\chi^2_{\nu}$ of the ASAS-SN light curve. While the majority of these trials do not affect the phasing of the light curve significantly, we find that there are some combinations of RV measurements that, when removed, produce a better phasing agreement between the light curve and the radial velocity measurements.

\begin{figure*}
    \centering
    \includegraphics[width=\linewidth]{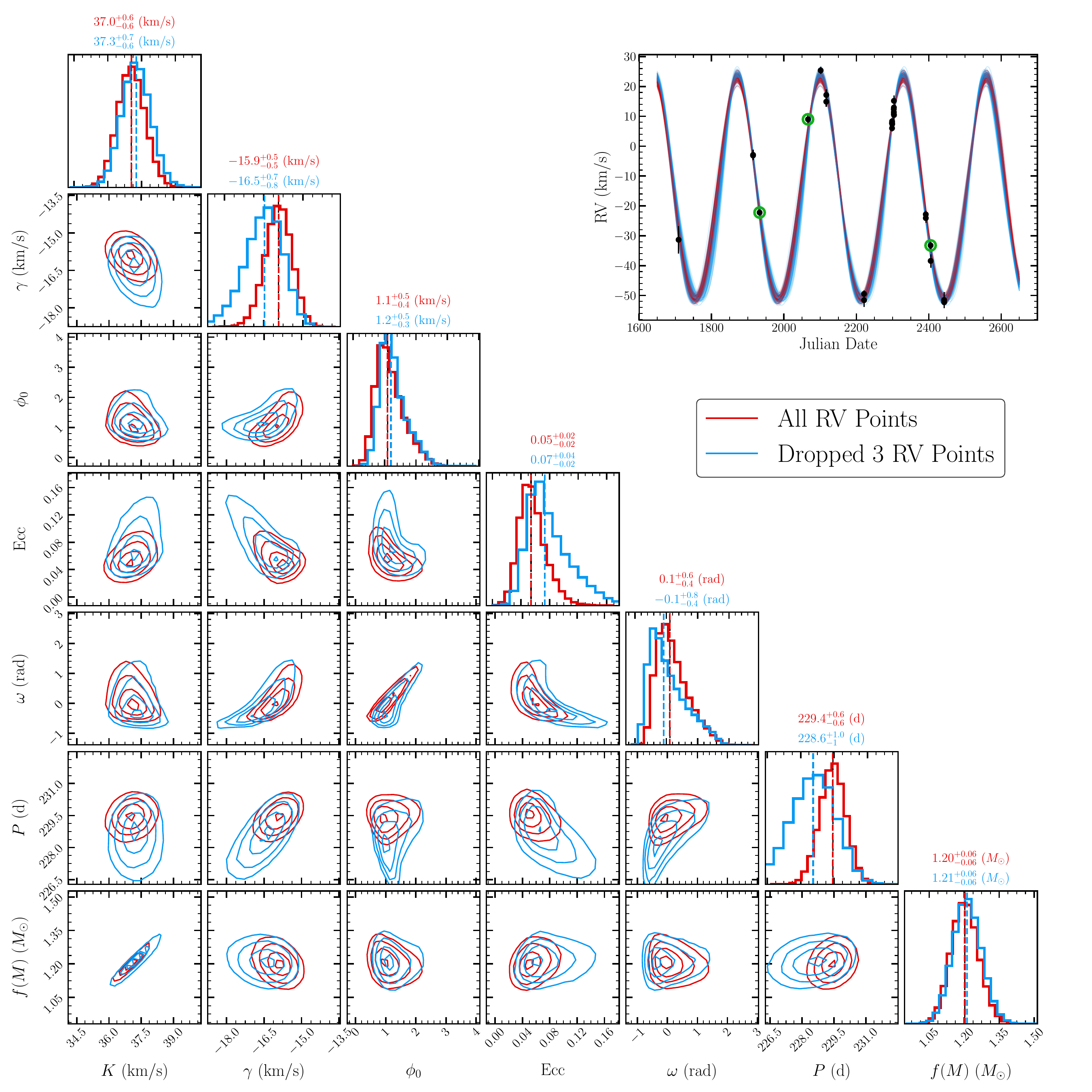}
    \caption{Comparison of the MCMC posteriors for the Keplerian orbit fit of J1640 using the full set of \Gaia{} RVs (red) and one with three points removed (blue). The posteriors are nearly identical, but the slight change in the orbital period is enough to improve the phasing of the ASAS-SN light curve compared to the expectation from ellipsoidal variability. The RV orbits are shown in the upper right and the three dropped measurements are marked in green.}
    \label{fig:dropout_example}
\end{figure*}

Figure \ref{fig:dropout_example} shows an example where three RV points have been removed. The RV fit looks nearly identical to the fit using all of the \Gaia{} RVs, and the MCMC posteriors show the results are consistent within $1\sigma$. In this case, the orbital period differs by $\sim 0.9$~days, which can produce an appreciable phase shift over the course of the five year gap between the \Gaia{} RVs and the ASAS-SN light curve.

Additional \PHOEBE{} models that exclude subsets of RV measurements could be used to better understand which RV measurements contribute most to the poor fit shown in Figure \ref{fig:rvmodel_comparison}. Ground-based radial velocity follow-up could also help further constrain the RV orbit of this target. An extended baseline could also place stronger constraints on the possibility of apsidal motion. Ultimately, however, the large RV residuals for the \PHOEBE{} model of J1640 do not alter the overall characterization of this system as a high mass function ellipsoidal variable on the giant branch.

\begin{table*}
    \centering
    \caption{\PHOEBE{} MCMC posteriors for J0946 and J1640 using the ASAS-SN light curves and \Gaia{} RVs. For each target we fit two models. The ``Dark Companion'' model assumes that the secondary is compact ($R_2=1\times10^{-6}\ R_\odot$) and dark ($T_{\rm{eff},2}=300\ \rm{K}$). The primary mass in these models is fixed to be the core mass using the SED fit results (Tables \ref{tab:sed_table_j0946} and \ref{tab:sed_table_j1640} and Equation \ref{eqn:coremass}). For the ``luminous comapnion'' models we do not restrict the secondary to be a compact object and use $M_1$ as a free parameter. The reduced $\chi^2$ of the light curve for each model is given by $\chi^2_{\nu, \rm{LC}}$.}
    \sisetup{table-auto-round,
     group-digits=false}
    \renewcommand{\arraystretch}{1.5} % Adjust the value as needed
    \setlength{\tabcolsep}{12pt}
    \begin{center}
        \input{ANC/phoebe_table}
    \end{center}
    \label{tab:phoebetable}
\end{table*}

\section{Discussion and Conclusions} \label{sec:discussion}

We use the \Gaia{} DR3 light curves and RVs from the recent \Gaia{} Focused Product Release \citep{Gaia2023_FPR} to identify eight targets with large binary mass functions, all of which are on the giant branch. The presence of a massive, unseen companion suggests that these systems are candidates for hosting a non-interacting compact object. 

We use the \Gaia{} and ASAS-SN light curves to search for eclipses (Figure \ref{fig:lightcurves}). For three targets, J0131, J1030, and J2059, we find eclipses consistent with the orbital period from the radial velocities, and rule these out as BH candidates. J1513 shows no apparent photometric variability in ASAS-SN, and the scatter in the \Gaia{} light curve makes interpretation challenging. We remove this system from consideration. 

We then fit the spectroscopic orbit models for the ELL systems (Figure \ref{fig:rv_orbits} and Table \ref{tab:rvorbits}). All are consistent with circular orbits and we find RV semiamplitudes comparable with the results from \citet{Gaia2023_FPR}. For the remaining four targets, we attempt to characterize the red giant primaries using broadband spectral energy distributions. Unfortunately, J0812 is blended with a nearby bright star. For the three other ELLs, J0946, J1640, and J2017, we fit the SED using \Gaia{}, 2MASS, and \WISE{} photometry (Figure \ref{fig:sed_panel}). J0946 and J1640 both have infrared excesses that are well described by a model including circumstellar dust. The dust in these systems could come from a radiatively driven wind, mass loss episodes, or vestiges of a previous period of binary interaction and mass transfer. There is some evidence of systematic issues with the \WISE{} photometry from diffraction spikes and blending, and additional IR observations could be used to better characterize the IR properties of these systems.

J2017 (V2012 Cyg, HD332077), previously classified as an S-star \citep{Jorissen92}, is one of the most luminous targets in the \Gaia{} FPR catalog. The radius from the SED fit and the orbital period require the photometric primary to be fairly massive $\gtrsim 10\ M_\odot$ in order for the giant to not overflow its Roche lobe. This characterization is at odds with the description from \citet{Jorissen92} that the binary is composed of a stripped giant with a main sequence A-star companion. A possible reconciliation is that the binary motion introduces a bias in the \Gaia{} parallax leading to an overestimation of the radius since $R\propto 1/\varpi$. Ultraviolet observations of this target with \textit{Swift} could be useful in identifying and characterizing a luminous secondary. Because of the evidence of a second component in the IUE spectrum \citep{Ake92}, we remove this target from consideration as a non-interacting compact object candidate.  

Finally, for J0946 and J1640 we use \PHOEBE{} to fit the ASAS-SN light curves and radial velocities (Figures \ref{fig:J0946_corner} and \ref{fig:J1640_corner}). Both J0946 and J1640 can be described by a stripped giant with a $\sim 3.7\ M_\odot$ companion. Based on the SED fit, a main sequence companion of this mass could be detected for both systems. However, it is also possible that these systems are analogs to the Giraffe and the Unicorn \citep{Jayasinghe21_unicorn, Jayasinghe22}, both of which have subgiant companions with a lower temperature more similar to the giant. Such a system would not stand out in the UV. 

We ran additional \PHOEBE{} models where the companion is treated as a luminous star rather than a compact object. In the case of J0946, the two-star \PHOEBE{} model predicts a binary with two giants and a luminosity ratio $L_2/L_1=0.85$. We would expect to detect this as a double-lined spectroscopic binary unless the companion is rapidly rotating and the spectral lines are sufficiently broadened. The two-star \PHOEBE{} model of J1640 finds a cool companion ($T_{\rm{eff, 2}}=3800^{+400}_{-200}$~K with $M_2=3.8\ M_\odot$ and radius $R_2=4^{+4}_{-2}\ R_\odot$, which is not consistent with single-star evolutionary models. Additional \PHOEBE{} models including IR light curves would improve the constraints on a luminous companion. The possible systematic problems with the \WISE{} observations limit the utility of the \WISE{} light curves.  

Unlike the two strongest non-interacting binary candidates, \Gaia{}-BH1 \citep{ElBadry23_BH1, Chakrabarti23} and \Gaia{}-BH2 \citep{Tanikawa23, ElBadry23_BH2}, both J0946 and J1640 have circular orbits and are close enough for tidal distortion to produce large amplitude ellipsoidal modulations. While the natal kick distribution is thought to be smaller for BHs than for neutron stars \citep{Repetto12, Mandel20}, we would expect these orbits to be moderately eccentric following the supernovae. Still, tidal effects would be expected to circularize the binaries within timescales short compared to the evolution of the giant. 

The circularization timescale is
\begin{equation}
    \tau_c = f\left(\frac{M_{\rm{env}}R_1^2}{L_1}\right)^{1/3} \frac{M_1^3}{M_{\rm{env}}M_2(M_1+M_2)}\left(\frac{a}{R}\right)^8,
\end{equation}
\noindent where $M_1$ is the mass of the giant primary, $M_{\rm{env}}$ is the mass of the envelope of the giant, and $M_2$ is the mass of the companion \citep{Verbunt95, Zahn77}. The quantity $f$ is a dimensionless factor of order unity that contains physics related to the convective and viscous properties of the star. For J0946, even if the primary mass is typical of a red giant $\sim 1\ M_\odot$ with $M_{\rm{env}} = M_1/2$, the binary would be expected to circularize within $\sim 75,000$~years for companions $\lesssim 10\ M_\odot$.

High-resolution spectra could be used to make more direct comparisons to V723 Mon \citep{Jayasinghe21_unicorn} and  2M04123153$+$6738486 \citep{Jayasinghe22}. \citet{ElBadry22_zoo} used spectral disentangling to identify companions and rule out compact object companions in the systems. The stripped red giants in these systems dominate the combined SED redder than $\sim 5000$~\AA, and identifying the subgiant companions is made more challenging due to their rotationally broadened spectral lines from being tidally locked. For J0946 and J1640, the high extinction could make obtaining spectra with high enough signal-to-noise spectra for disentangling challenging. A subgiant companion would be expected to dominate the SED at $\lambda \lesssim 5000$~\AA. 

In the SDSS $u$-band (effective wavelength $\lambda=3608$~\AA), the SED models of J0946 and J1640 predict extinction-corrected magnitudes for the giants of $u\sim17.9$~mag and $u\sim19.3$~mag respectively. Even if a subgiant companion contributed four times as much flux at this wavelength, the magnitude of the binary would still be fainter than $16.1$~mag and $17.6$~mag for J0946 and J1640, respectively. Multi-band \textit{Swift} UVOT or Hubble Space Telescope UV photometry may be a more promising method to detect any subgiant companions hotter than the giant stars. 

Searching for and characterizing non-interacting BHs requires careful consideration of false positives originating from mass transfer. Although \Gaia{} DR3 was predicted to detect dozens to hundreds of BHs \citep[e.g.,][]{Breivik17, Chawla23}, only two strong candidates have been detected. While this could indicate a need for revision of binary population synthesis models, both of the \Gaia{} astrometric BHs are within 1.2~kpc, suggesting similar systems almost certainly exist all over the Galaxy. Spectroscopic and astrometric surveys are the most promising pathways to identify new candidates. \Gaia{} DR3 included spectroscopic orbit fit results for $\sim 185,000$ binaries and astrometric orbit fits for $\sim 130,000$ binaries. The next \Gaia{} data release is expected to include time-series RVs and astrometric measurements for a larger sample of systems. This \Gaia{} Focused Product Release includes the first \Gaia{} epoch RVs for binary systems and serves as a demonstration of the potential for future \Gaia{} data releases to aid in the search for non-interacting black holes. 

\section*{Acknowledgements}

We thank the anonymous referee for their comments. We thank Michael Tucker for his help in deriving upper limits with \textit{GALEX} and David Martin, Christine Daher, and Kareem El-Badry for useful discussions. We thank Las Cumbres Observatory and its staff for their continued support of ASAS-SN. ASAS-SN is funded in part by the Gordon and Betty Moore Foundation through grants GBMF5490 and GBMF10501 to the Ohio State University, and also funded in part by the Alfred P. Sloan Foundation grant G-2021-14192. 

KZS and CSK are supported by NSF grants AST-1907570 and 2307385.

This work presents results from the European Space Agency space mission Gaia. Gaia data are being processed by the Gaia Data Processing and Analysis Consortium (DPAC). Funding for the DPAC is provided by national institutions, in particular the institutions participating in the Gaia MultiLateral Agreement.

\bibliographystyle{mnras}
\bibliography{gaiafpr} % if your bibtex file is called example.bib

\appendix

\section{\TESS{} Light Curve of J0131} \label{appendix:tess}

\begin{figure}
    \centering
    \includegraphics[width=\linewidth, trim={0 2.5cm 0cm 0cm},clip]{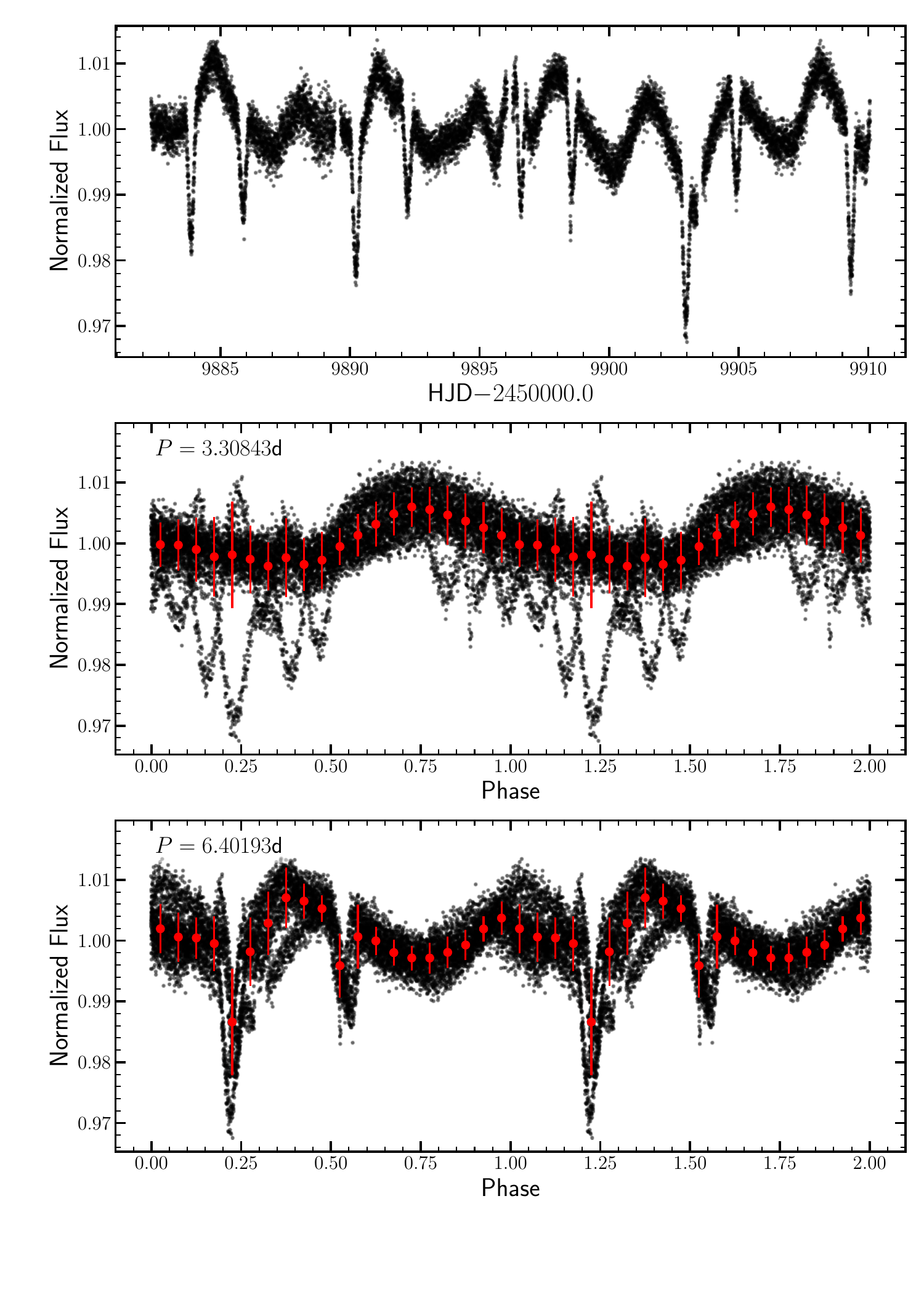}
    \caption{\textit{TESS} Sector 58 light curve of J0131 downloaded from the SPOC pipeline using the PDCSAP flux. The top panels shows the unfolded light curve. The middle and bottom panels show the light curve folded at the periods from the LS and BLS periodograms, respectively. The red points show the binned light curve.}
    \label{fig:j0131_tess}
\end{figure}

In Section \S\ref{sec:eclipsingbinaries} we used the QLP, SPOC, and TGLC pipelines to download \TESS{} light curves for the eight high $f(M)$ targets in Table \ref{tab:target_selection}. J0131 shows evidence for additional eclipses in the Sector 58 SPOC light curve shown in Figure \ref{fig:j0131_tess}. We use a Lomb-Scargle (LS) periodogram \citep{Lomb76, Scargle82} to identify a signal at $P_{\rm{LS}}=3.30843$ and a box-least-squares (BLS) periodogram \citep{Kovacs02} to identify a second $P_{\rm{BLS}}=6.40193$. Figure \ref{fig:j0131_tess} shows the \TESS{} light curve folded at both periods. 

Neither of these signals are present in the QLP light curves. The Sector 24 and 25 TGLC light curves show some evidence for additional variability, but it is noisier than the SPOC Sector 58 light curve. No other Sectors are available from the SPOC pipeline. The large \TESS{} pixels mean that this variable signal could be produced by a nearby variable star. Following \citet{Rowan23}, we cross-match with variable star catalogs to search for the contaminant with a 5' radius from the \Gaia{} position. There are 5 variable stars in the Zwicky Transient Facility variable star catalog \citep{Chen20} within 5'. One target, J013146.49+621953.5 is $95\farcs1$ away from J0131 and \citet{Chen20} reports a period $P_{\rm{ZTF}} = 3.36842$. Since $P_{\rm{ZTF}}\approx P_{\rm{LS}}$, and the $g$-band light curve shape of J013146.49+621953.5 is similar to the middle panel of Figure \ref{fig:j0131_tess}, the variability seen in the \TESS{} light curve of J0131 is likely due to contamination. 

The ZTF light curve of J013146.49+621953.5 does not show evidence for the $P_{\rm{BLS}}=6.40193$ eclipsing signal, and none of the other four nearby ZTF variables have a similar period. We find no matches to the ASAS-SN \citep{Jayasinghe21, Christy23} or ATLAS \citep{Heinze18} variable star catalogs. Although it is possible that this signal is intrinsic to the J0131, making it a triply-eclipsing binary, since the signal is only present in some \TESS{} Sectors and with some reduction pipelines, it seems more likely to be contamination from a nearby blended variable star.

\label{lastpage}

\end{document}

%% file: ANC/target_selection_table.tex
\begin{tabular}{l l S[table-format=3.4] S[table-format=2.4] S[table-format=2.1] S[table-format=2.3] S[table-format=3.1] S[table-format=2.1] S[table-format=1.2] r c}
\toprule
           {Source} & {Name} &       {RA} &      {DEC} &     {$G$} &  {RUWE} &      {$P$} &     {$K$} &  {$f(M)$} &            {Distance} &  {ELL} \\
{} & {} & {(deg)} & {(deg)} & {(mag)} & {} & {(day)} & {(km/s)} & {($M_\odot$)} & {(pc)} & {}\\ 
\midrule
 512307478642441984 &  J0131 &  22.986343 &  62.349056 & 11.071171 &   1.274 & 145.781827 & 59.194690 &  3.133097 &  $3600^{+200}_{-300}$ & \xmark \\
5541400855806539392 &  J0812 & 123.208510 & -37.720467 & 12.323505 &   1.065 & 680.404462 & 29.751371 &  1.856566 &  $5300^{+600}_{-500}$ & \cmark \\
5405789050140488320 &  J0946 & 146.722062 & -51.201720 & 13.239762 &   1.037 & 102.419502 & 54.841510 &  1.750383 &  $4300^{+200}_{-200}$ & \cmark \\
5351477646814401408 &  J1030 & 157.563293 & -58.378805 & 12.933435 &   0.987 & 113.299287 & 46.498688 &  1.180243 &  $3800^{+300}_{-200}$ & \xmark \\
5887977370590405632 &  J1513 & 228.356804 & -53.606859 & 13.800054 &   1.265 & 788.442771 & 24.898160 &  1.260939 & $5600^{+1100}_{-700}$ & \xmark \\
5968984160290449024 &  J1640 & 250.097968 & -41.269346 & 13.227487 &   1.559 & 229.798565 & 36.612880 &  1.168614 &  $3800^{+400}_{-400}$ & \cmark \\
2053893497434322048 &  J2017 & 304.296061 &  31.555016 &  7.734626 &   2.262 & 670.724647 & 26.406700 &  1.279702 &  $1700^{+200}_{-200}$ & \cmark \\
2162167694508896128 &  J2059 & 314.809215 &  43.629021 & 10.637851 &   1.181 & 289.568551 & 34.828793 &  1.267619 &  $2500^{+100}_{-100}$ & \xmark \\
\bottomrule
\end{tabular}

%% file: ANC/rvorbits_table.tex
\begin{tabular}{l r r r r r r r r}
\toprule
{} &                      J0812 &                     J0946 &                   J1513 &                   J1640 &                      J2017 \\
\midrule
{Period (d)}                         &            $680^{+2}_{-2}$ &  $102.60^{+0.07}_{-0.08}$ &       $810^{+30}_{-30}$ &   $229.4^{+0.6}_{-0.6}$ &            $670^{+1}_{-1}$ \\
{$K$ (km/s)}                         &       $29.8^{+0.3}_{-0.3}$ &            $56^{+1}_{-1}$ &          $24^{+2}_{-2}$ &    $37.0^{+0.6}_{-0.6}$ &       $26.5^{+0.1}_{-0.1}$ \\
{$e$ }                               &  $0.076^{+0.008}_{-0.008}$ &    $0.07^{+0.02}_{-0.02}$ &  $0.13^{+0.07}_{-0.10}$ &  $0.05^{+0.02}_{-0.02}$ &  $0.029^{+0.005}_{-0.005}$ \\
{$f(M)$ ($M_\odot$)}                 &     $1.86^{+0.05}_{-0.05}$ &       $1.9^{+0.1}_{-0.1}$ &     $1.1^{+0.3}_{-0.3}$ &  $1.20^{+0.06}_{-0.06}$ &     $1.29^{+0.02}_{-0.02}$ \\
{$\gamma$ (km/s)}                    &      $123.5^{+0.2}_{-0.2}$ &      $25.0^{+0.6}_{-0.6}$ &         $-52^{+1}_{-2}$ &   $-15.9^{+0.5}_{-0.5}$ &    $-6.38^{+0.09}_{-0.09}$ \\
{$t_0$ }                             &        $2456329^{+9}_{-8}$ &       $2456866^{+1}_{-1}$ &   $2456720^{+50}_{-60}$ &     $2456864^{+6}_{-6}$ &        $2456604^{+4}_{-4}$ \\
{$\omega$ (rad)}                     &     $3.03^{+0.09}_{-0.07}$ &      $-1.7^{+0.3}_{-0.2}$ &    $-2.2^{+0.4}_{-0.5}$ &     $0.1^{+0.4}_{-0.6}$ &        $1.4^{+0.2}_{-0.2}$ \\
{$\frac{\rm{orb\ motion}}{\varpi}$ } &     $1.85^{+0.02}_{-0.02}$ &    $0.52^{+0.01}_{-0.01}$ &     $1.8^{+0.2}_{-0.1}$ &  $0.78^{+0.01}_{-0.01}$ &  $1.624^{+0.009}_{-0.009}$ \\
\bottomrule
\end{tabular}

%% file: ANC/J0946/revised_sed_table_fullJ0946.tex
\begin{tabular}{cc r r r r r r r r r r r r r r}
\toprule
{Model} &                    Free &                 Fixed &                Binary &                    Dust \\
\midrule
{$T_{\rm{eff}}$ $(\rm{K})$}    &      $4150^{+50}_{-40}$ &    $4140^{+10}_{-10}$ &    $3980^{+30}_{-50}$ &    $4410^{+120}_{-100}$ \\
{$R$ $(\rm{R}_\odot)$}         &          $33^{+1}_{-1}$ &        $33^{+1}_{-1}$ &  $34.2^{+0.6}_{-0.7}$ &          $28^{+1}_{-1}$ \\
{$L$ $(\rm{L}_\odot)$}         &       $300^{+30}_{-30}$ &     $290^{+20}_{-20}$ &     $260^{+10}_{-20}$ &        $265^{+9}_{-16}$ \\
{$\log g$ }                    &     $2.8^{+0.2}_{-0.2}$ &   $2.8^{+0.2}_{-0.2}$ &   $1.6^{+0.2}_{-0.2}$ &                      -- \\
{$T_{\rm{eff, 2}}$ $(\rm{K})$} &                      -- &                    -- &  $6200^{+800}_{-700}$ &                      -- \\
{$R_2$ $(\rm{R}_\odot)$}       &                      -- &                    -- &         $5^{+3}_{-2}$ &                      -- \\
{$L_2$ $(\rm{L}_\odot)$}       &                      -- &                    -- &       $38^{+15}_{-9}$ &                      -- \\
{$\log g_2$ }                  &                      -- &                    -- &         $3^{+2}_{-2}$ &                      -- \\
{$A_V$ $(\rm{mag})$}           &  $1.51^{+0.07}_{-0.07}$ &          \textbf{1.5} &          \textbf{1.5} &   $1.39^{+0.05}_{-0.2}$ \\
{Dist $(\rm{pc})$}             &    $4400^{+200}_{-200}$ &  $4400^{+200}_{-200}$ &         \textbf{4297} &                      -- \\
{$\tau_{\nu}$ }                &                      -- &                    -- &                    -- &  $0.22^{+0.05}_{-0.06}$ \\
{$T_{\rm{dust}}$ $(\rm{K})$}   &                      -- &                    -- &                    -- &       $840^{+70}_{-70}$ \\
{$R_{\rm{dust}}$ $(R_\odot)$}  &                      -- &                    -- &                    -- &            $1400\pm300$ \\
{$M_c$ $(M_\odot)$}            &                   0.329 &                 0.328 &                 0.323 &                   0.324 \\
\bottomrule
\end{tabular}

%% file: ANC/J1640/revised_sed_table_fullJ1640.tex
\begin{tabular}{cc r r r r r r r r r r r r r r}
\toprule
{Model} &                    Free &                  Fixed &                  Binary &                    Dust \\
\midrule
{$T_{\rm{eff}}$ $(\rm{K})$}    &      $4080^{+50}_{-40}$ &     $3630^{+20}_{-20}$ &    $3300^{+100}_{-100}$ &      $3910^{+50}_{-50}$ \\
{$R$ $(\rm{R}_\odot)$}         &          $56^{+5}_{-5}$ &         $59^{+4}_{-5}$ &    $63.2^{+0.3}_{-0.3}$ &          $46^{+1}_{-1}$ \\
{$L$ $(\rm{L}_\odot)$}         &     $800^{+100}_{-100}$ &      $540^{+80}_{-80}$ &       $440^{+70}_{-60}$ &       $440^{+20}_{-20}$ \\
{$\log g$ }                    &     $3.0^{+0.1}_{-0.2}$ &  $1.58^{+0.1}_{-0.09}$ &  $0.04^{+0.05}_{-0.03}$ &                      -- \\
{$T_{\rm{eff, 2}}$ $(\rm{K})$} &                      -- &                     -- &    $7800^{+100}_{-300}$ &                      -- \\
{$R_2$ $(\rm{R}_\odot)$}       &                      -- &                     -- &     $3.7^{+0.3}_{-0.2}$ &                      -- \\
{$L_2$ $(\rm{L}_\odot)$}       &                      -- &                     -- &          $45^{+3}_{-2}$ &                      -- \\
{$\log g_2$ }                  &                      -- &                     -- &     $2.9^{+0.9}_{-0.9}$ &                      -- \\
{$A_V$ $(\rm{mag})$}           &  $3.39^{+0.08}_{-0.08}$ &          \textbf{2.44} &           \textbf{2.44} &   $1.77^{+0.2}_{-0.10}$ \\
{Dist $(\rm{pc})$}             &    $3800^{+300}_{-400}$ &   $4000^{+300}_{-300}$ &           \textbf{3809} &                      -- \\
{$\tau_{\nu}$ }                &                      -- &                     -- &                      -- &  $0.68^{+0.09}_{-0.08}$ \\
{$T_{\rm{dust}}$ $(\rm{K})$}   &                      -- &                     -- &                      -- &       $630^{+30}_{-20}$ \\
{$R_{\rm{dust}}$ $(R_\odot)$}  &                      -- &                     -- &                      -- &            $4000\pm500$ \\
{$M_c$ $(M_\odot)$}            &                   0.377 &                  0.358 &                   0.347 &                   0.348 \\
\bottomrule
\end{tabular}

%% file: ANC/J2017/sed_table_fullJ2017.tex
\begin{tabular}{cc r r r r r r r r r r r}
\toprule
{Model} &                     Free &                    Fixed &                  Binary \\
\midrule
{$T_{\rm{eff}}$ $(\rm{K})$}    &      $3420^{+70}_{-150}$ &      $3410^{+70}_{-150}$ &       $3506^{+41}_{-5}$ \\
{$R$ $(\rm{R}_\odot)$}         &        $390^{+30}_{-40}$ &        $330^{+30}_{-30}$ &      $220^{+70}_{-150}$ \\
{$L$ $(\rm{L}_\odot)$}         &  $18000^{+4000}_{-4000}$ &  $13000^{+3000}_{-3000}$ &  $6000^{+5000}_{-6000}$ \\
{$\log g$ }                    &      $1.1^{+0.1}_{-0.2}$ &   $0.09^{+0.07}_{-0.05}$ &     $0.08^{+2}_{-0.07}$ \\
{$T_{\rm{eff, 2}}$ $(\rm{K})$} &                       -- &                       -- &    $3300^{+200}_{-200}$ \\
{$R_2$ $(\rm{R}_\odot)$}       &                       -- &                       -- &      $220^{+80}_{-130}$ \\
{$L_2$ $(\rm{L}_\odot)$}       &                       -- &                       -- &  $5000^{+4000}_{-4000}$ \\
{$\log g_2$ }                  &                       -- &                       -- &     $0.12^{+2}_{-0.10}$ \\
{$A_V$ $(\rm{mag})$}           &   $2.44^{+0.07}_{-0.08}$ &            \textbf{1.71} &           \textbf{1.71} \\
{Dist $(\rm{pc})$}             &     $1900^{+200}_{-200}$ &     $1900^{+200}_{-200}$ &           \textbf{1729} \\
{$M_c$ $(M_\odot)$}            &                     0.83 &                     0.74 &                    0.63 \\
\bottomrule
\end{tabular}

%% file: ANC/phoebe_table.tex
\begin{tabular}{ccccc}
\toprule
\multicolumn{1}{c}{} & \multicolumn{2}{c}{J0946} & \multicolumn{2}{c}{J1640} \\
\toprule
{} &              Dark Companion &         Luminous Companion &             Dark Companion &         Luminous Companion \\
\midrule
Period (d)                &  $102.524^{+0.01}_{-0.009}$ &   $102.53^{+0.01}_{-0.01}$ &   $228.13^{+0.04}_{-0.04}$ &   $228.14^{+0.04}_{-0.04}$ \\
$t_0$                     &   $2458252.5^{+0.1}_{-0.1}$ &  $2458252.4^{+0.1}_{-0.1}$ &  $2458271.9^{+0.3}_{-0.3}$ &  $2458271.6^{+0.3}_{-0.3}$ \\
$e$                       &   $0.070^{+0.004}_{-0.004}$ &  $0.064^{+0.004}_{-0.004}$ &  $0.014^{+0.002}_{-0.003}$ &  $0.031^{+0.002}_{-0.002}$ \\
$q$                       &        $11.8^{+0.8}_{-0.7}$ &             $10^{+2}_{-2}$ &       $10.5^{+0.9}_{-0.6}$ &              $9^{+1}_{-1}$ \\
Incl $(^{\circ})$         &              $56^{+2}_{-2}$ &             $56^{+2}_{-2}$ &         $50.0^{+1.0}_{-2}$ &             $51^{+1}_{-1}$ \\
$R_1\ (R_\odot)$          &        $20.4^{+0.3}_{-0.3}$ &             $28^{+1}_{-1}$ &       $45.6^{+0.7}_{-0.3}$ &             $47^{+2}_{-2}$ \\
$T_{\rm{eff},1} (\rm{K})$ &         $4510^{+50}_{-120}$ &        $4430^{+60}_{-150}$ &         $3830^{+40}_{-30}$ &         $3900^{+40}_{-50}$ \\
$R_2\ (R_\odot)$          &            $1\times10^{-6}$ &             $22^{+3}_{-2}$ &           $1\times10^{-6}$ &              $4^{+4}_{-2}$ \\
$T_{\rm{eff},2} (\rm{K})$ &                       300.0 &       $4800^{+200}_{-200}$ &                      300.0 &       $3800^{+400}_{-200}$ \\
$\gamma\ (\rm{km/s})$     &        $25.2^{+0.6}_{-0.6}$ &       $25.3^{+0.6}_{-0.6}$ &      $-16.1^{+0.3}_{-0.3}$ &      $-15.8^{+0.3}_{-0.3}$ \\
$M_1\ (M_\odot)$          &                        0.32 &     $0.42^{+0.07}_{-0.06}$ &                       0.35 &     $0.39^{+0.05}_{-0.04}$ \\
$M_2\ (M_\odot)$          &         $3.8^{+0.3}_{-0.2}$ &        $3.9^{+0.2}_{-0.2}$ &        $3.7^{+0.3}_{-0.2}$ &        $3.7^{+0.2}_{-0.2}$ \\
$\chi^2_{\nu,\ \rm{LC}}$  &                        5.36 &                       5.41 &                      16.15 &                      15.97 \\
\bottomrule
\end{tabular}